%% file: main.tex
\RequirePackage{rotating}
\RequirePackage{fix-cm}

\documentclass[final,natbib]{svjour3}       
\smartqed  
\usepackage{amsmath,amssymb,amsfonts}
\usepackage{algorithmic}
\usepackage{graphicx}
\usepackage{textcomp}
\usepackage{xcolor}
\usepackage[T1]{fontenc}
\usepackage[skins]{tcolorbox}

\journalname{Empirical Software Engineering}
\input{preamble}

\begin{document}
	
	\title{What are the Machine Learning best practices reported by practitioners on Stack Exchange?
	}
	\titlerunning{What are the ML best practices reported by practitioners on Stack Exchange?}

	\author{Anamaria Mojica-Hanke         \and
		Andrea Bayona \and  Mario Linares-Vásquez \and
		Steffen Herbold \and
		Fabio A. González
	}
	
	
	\institute{Anamaria Mojica-Hanke \at
		Universidad de los Andes, Colombia.
		\email{ai.mojica10@uniandes.edu.co}   
		\and 
		Andrea Bayona \at
		Universidad de los Andes, Colombia.
		\email{ma.bayona@uniandes.edu.co}   
		\and 
		Mario Linares-Vásquez  \at
		Universidad de los Andes, Colombia.
		\email{m.linaresv@uniandes.edu.co}   
		\and 
		Steffen Herbold\at
		Technische Universität Clausthal, Germany.
		\email{steffen.herbold@tu-clausthal.de}   
		\and 
		Fabio A. González\at
		Universidad Nacional, Colombia.
		\email{fagonzalezo@unal.edu.co}   
	}
	
	\date{Received: date / Accepted: date}

	\maketitle
	
	\begin{abstract}
	
 \acf{ML}  is being used in multiple disciplines due to its powerful capability to infer relationships within data. In particular, \acf{SE}  is one of those disciplines in which \ac{ML} has been used for multiple tasks, like software categorization, bugs prediction, and testing. In addition to the multiple \ac{ML} applications, some studies have been conducted to detect and understand possible pitfalls and issues when using \ac{ML}. However, to the best of our knowledge, only a few studies have focused on presenting \ac{ML} best practices or guidelines for the application of  \ac{ML} in different domains. In addition,  the practices and literature presented in previous literature (i) are domain-specific (\eg concrete practices in biomechanics), (ii)  describe few practices, or (iii) the practices lack rigorous validation and are presented in gray literature. 
 In this paper, we present a study listing 127 \ac{ML} best practices systematically mining 242 posts of 14 different \acf{STE} websites and validated by four independent \ac{ML} experts. The list of practices is presented in a set of categories related to different stages of the implementation process of an \ac{ML}-enabled system; for each practice, we include explanations and examples. In all the practices, the provided examples focus on \ac{SE} tasks. We expect this list of practices could help  practitioners to understand better the practices and use ML in a more informed way, in particular newcomers to this new area  that sits at the intersection of software engineering and machine learning.


	\end{abstract}
	
	\input{section_introduction}

\input{section_related_work}

	\input{section_design_study}

	\input{section_result}\input{section_discussion}
	\input{section_threats}

	\input{section_conclusion}

	\begin{acknowledgements}
		
		We would like to thank   David Koll, Grace A. Lewis, Fabian Kortum,  Haydemar Castro Nuñez,  Monica Gutierrez,   and  Valerie Parra for their reviews and collaboration in the study and manuscript.  

	\end{acknowledgements}

	%
	%

\bibliographystyle{spbasic}      
\bibliography{bib_local,bib_tools}   
	
\end{document}

%% file: preamble.tex
\usepackage{rotating}
\usepackage{adjustbox}
\usepackage{amssymb}
\usepackage{natbib}
\usepackage{etoolbox}
\usepackage{longtable}
\usepackage{array,booktabs}
\newcommand{\tabitem}{~~\llap{-}~~}
\newcolumntype{M}[1]{>{\centering\arraybackslash}m{#1}}
\usepackage{float}
\usepackage[T1]{fontenc}

\usepackage{lmodern}

\makeatletter

\patchcmd{\NAT@citex}
{\@citea\NAT@hyper@{%
		\NAT@nmfmt{\NAT@nm}%
		\hyper@natlinkbreak{\NAT@aysep\NAT@spacechar}{\@citeb\@extra@b@citeb}%
		\NAT@date}}
{\@citea\NAT@nmfmt{\NAT@nm}%
	\NAT@aysep\NAT@spacechar\NAT@hyper@{\NAT@date}}{}{}

\patchcmd{\NAT@citex}
{\@citea\NAT@hyper@{%
		\NAT@nmfmt{\NAT@nm}%
		\hyper@natlinkbreak{\NAT@spacechar\NAT@@open\if*#1*\else#1\NAT@spacechar\fi}%
		{\@citeb\@extra@b@citeb}%
		\NAT@date}}
{\@citea\NAT@nmfmt{\NAT@nm}%
	\NAT@spacechar\NAT@@open\if*#1*\else#1\NAT@spacechar\fi\NAT@hyper@{\NAT@date}}
{}{}

\makeatother

\usepackage[plain]{fancyref}
\usepackage{ifdraft}

\usepackage[inline]{enumitem}
\usepackage{xcolor}
\usepackage{xspace}
\usepackage[final]{listings}
\usepackage{acronym}
\usepackage{hyperref}
\usepackage{url}
\usepackage{amsmath}
\usepackage{amssymb}
\usepackage{pifont}
\newcommand{\xmark}{\tiny{\ding{53}}}%
\usepackage{booktabs} 
\usepackage{subfig}
\usepackage{balance}
\usepackage{dirtree}
\usepackage{xstring}
\usepackage{ifthen}
\usepackage{mfirstuc}
\usepackage{enumitem}
\usepackage{amssymb}
\usepackage{multirow}
\usepackage[ruled]{algorithm2e} 

\hypersetup{%
	pdfborder = {0 0 0},
	colorlinks,
	citecolor=blue,
	filecolor=blue,
	linkcolor=blue,
	urlcolor=blue
}

\usepackage{etoolbox}
\makeatletter
\patchcmd{\@makecaption}
  {\scshape}
  {}
  {}
  {}
\patchcmd{\@makecaption}
  {\\}
  {.\ }
  {}
  {}
\makeatother

\newcommand{\eg}{\emph{e.g.,}\xspace}

\newcommand{\ie}{\emph{i.e.,}\xspace}
\renewcommand{\etal}{\emph{et al.}\xspace}

\newcommand{\figref}[1]{Fig.~\ref{#1}\xspace}

\newcommand{\tabref}[1]{Table~\ref{#1}\xspace}

\definecolor{author}{rgb}{.5, .5, .5}
\definecolor{note}{rgb}{.54, .17, .89}
\definecolor{idea}{rgb}{.1, .7, .0}
\definecolor{missing}{rgb}{1, .03, .0}
\definecolor{deleteme}{rgb}{.9, .3, .5}
\definecolor{comment}{rgb}{0.19, 0.55, 0.91}

\newcommand{\MARIO}[2][comment]{\authorcomment[#1]{MLV}{#2}}
\newcommand{\STEFFEN}[2][comment]{\authorcomment[#1]{STH}{#2}}
\newcommand{\ANA}[2][comment]{\authorcomment[#1]{AMH}{#2}}

\newcommand{\authorcomment}[3][comment]
{\noindent
	\fbox{\footnotesize\textcolor{author}{\textsc{#2}}}
	\textcolor{#1}{\textsl{#3}}{}}

\input{acronyms}

\definecolor{graypr}{rgb}{0, 0, 0}
\newcommand{\practice}[1]{%
		{%
			\emph{#1}}}

\makeatletter
\newcommand{\setword}[3][]{%
	\ifthenelse{ \equal{#1}{}}%
	{\ensuremath{\mathbf{#2_{#3}}}\phantomsection\def\@currentlabel{\unexpanded{\ensuremath{\mathbf{\capitalisewords{#2}_{#3}}}}}\label{pr:#2#3}}%
	{\ensuremath{\mathbf{\capitalisewords{#2}_{#3}}}\phantomsection\def\@currentlabel{\unexpanded{\ensuremath{\mathbf{\capitalisewords{#2}_{#3}}}}}\label{pr:#2#3}}%
}
\makeatother

\makeatletter
\newcommand{\acron}[3][]{%
	\ifthenelse{ \equal{#1}{}}%
	{\ensuremath{\mathbf{#2_{#3}}}\phantomsection\def\@currentlabel{\unexpanded{\ensuremath{\mathbf{\capitalisewords{#2}_{#3}}}}}\label{pr:#2#3}}%
	{\ensuremath{\mathbf{\capitalisewords{#2}_{#3}}}\phantomsection\def\@currentlabel{\unexpanded{\ensuremath{\mathbf{\capitalisewords{#2}_{#3}}}}}\label{pr:#2#3}}%
}
\makeatother

\newcommand{\acronref}[3][]{%
	\ifthenelse{ \equal{#1}{}}%
	{\ensuremath{\mathbf{#2_{#3}}}\ref{pr:#2#3}}%
	{\ensuremath{\mathbf{\capitalisewords{#2}_{#3}}}\ref{pr:#2#3}}%
}

%% file: acronyms.tex
\acrodef{ML}{Machine Learning}
\acrodef{AI}{Artificial Intelligence}
\acrodef{SO}{Stack Overflow}
\acrodef{STE}{Stack Exchange}
\acrodef{SE}{Software Engineering}
\acrodef{CQA}{Community Question Answering}
\acrodef{CQ+A}[CQ\&A]{Community Question Answering}
\acrodef{LDA}{Latent Dirichlet Allocation}
\acrodef{DL}{Deep Learning}
\acrodef{MySQL}{local database }


%% file: section_introduction.tex

\section{Introduction}
\label{sec:intro}


 \acf{ML} methods have been applied in multiple domains like medicine \citep[\eg][]{Kamalov2022, rajkomar2019machine}, finance \citep[\eg][]{dixon2020machine, sarlin2017machine}, environmental science \cite[\eg][]{GIBERT20181}, and \ac{SE}, among others. In the latter, \ac{ML} has been used  for supporting different aspects and tasks related to software engineering, such as type inference in bug/vulnerability detection \citep[\eg][]{dinella2020hoppity, pradel2018deepbugs}, code completion \citep[\eg][]{kim2021code, svyatkovskiy2019pythia}; code summarization  \citep[\eg][]{wan2018improving, allamanis2016convolutional}; code representation/embedding \citep[\eg][]{niu2022spt, kanade2020learning}, code transformations \citep[\eg][]{li2020dlfix, tufano2019learning}, program translation  \citep[\eg][]{ahmad2021, roziere2020unsupervised}, empirical studies  \citep[\eg][]{zhou2021assessing, mazuera2021shallow, paltenghi2021thinking, chirkova2021empirical}, software categorization \citep[\eg][]{escobar2015unsupervised, linares2014using}; bug triage \citep[\eg][]{linares2012triaging, murphy2004automatic}.  For a more detailed description of how ML has been used for SE, we refer to the interested reader to \cite{Watson2022, devanbu2020deep,zhang2003machine}.

All in all, in the past years, we have seen wide adoption and usage of \ac{ML} in \ac{SE} and other disciplines; according to Gartner, this topic was at the ``peak of inflated expectations'' of the hype cycle during  2016\footnote{https://www.gartner.com/en/documents/3388917}\!, 2017\footnote{https://www.gartner.com/en/documents/3772081}\!, 2018\footnote{https://www.gartner.com/en/documents/3883664}\!, and 2019\footnote{https://www.gartner.com/smarterwithgartner/top-trends-on-the-gartner-hype-cycle-for-artificial-intelligence-2019}. However, in 2020\footnote{https://www.gartner.com/smarterwithgartner/2-megatrends-dominate-the-gartner-hype-cycle-for-artificial-intelligence-2020} and 2021\footnote{https://www.gartner.com/en/articles/the-4-trends-that-prevail-on-the-gartner-hype-cycle-for-ai-2021}\!, it has dropped down to the ``trough of disillusionment''. This transition  suggests that the interest in \ac{ML} has waned as experiments and implementations fail to deliver.  This reduction of interest for practitioners and researchers could be a consequence of existing  challenges and pitfalls encountered when applying and researching \ac{ML}. For example,  pitfalls have been reported when ML is applied in different domains such as financial forecasting  \citep{wasserbacher2021machine}, autism diagnostics \citep{bone2015applying}, when using omic data \citep{teschendorff2019avoiding}, genomics \citep{whalen2021navigating},  human movement biomechanics \citep{halilaj2018machine},   stock returns forecasting \citep{leung2021promises},  other \ac{ML}-enabled systems \citep{Vento2019, Clemmedsson2018, LewisGrace2021WAIN} and  in \ac{ML} research \citep[\eg][]{biderman2020pitfalls, MichaelLones2021}.

In order to prevent or at least mitigate some pitfalls when applying  \ac{ML}, best practices/guidelines have been proposed by \cite{halilaj2018machine}, \cite{ biderman2020pitfalls}, and \cite{ MichaelLones2021}, based on  their own experience and focused on their respective discipline. Note that there is a plethora of publications (\eg, books, blogs, papers, etc.) on the field of ML; for example, grey literature such as the article by \cite{zinkevich_2021} are publicly available and could be considered as a first step with  general practices derived from anecdotal evidence.   However, to the best of our knowledge, there are no handbooks listing best practices for using ML in specific fields. As of today there is no publication that presents a list of \ac{ML} pitfalls, or best practices  focused on SE practitioners or researchers, \ie software engineers and researchers using ML as part of their solutions or studies. As \ac{ML} is becoming more and more involved in \ac{SE} development projects, 
bad practices should be avoided to prevent inadequate model planning, implementation, tuning, testing, deployment, and monitoring of ML implementations.

In this paper, we present a handbook of ML practices that could be used by SE researchers and practitioners. To achieve this, we relied on the knowledge reported by developers and practitioners on 14 \ac{CQ+A} websites.  In particular, we analyzed 242  posts (with a score greater than 0) with the purpose of identifying \ac{ML} best practices reported by \acf{STE} contributors. Afterward, the identified practices were validated by four \ac{ML} experts. After the experts' validation, we obtained  a list of 127 practices that are grouped into 10 high-level categories. We discuss each practice and provide examples and references supporting the practices.  In addition, we have created a detailed online appendix with  (i) a taxonomy grouping the  127 practices, (ii) a table containing all the  practices with their associated \ac{STE} links, (iii)  and detailed explanations of the practices (\emph{\textbf{\url{https://tinyurl.com/bestPML}}}). We expect this paper and our findings would help SE researchers and practitioners to promote the usage of \ac{ML} best practices  in \ac{SE}, particularly for newcomers that could be overwhelmed by the plethora of existing publications.  

This paper is organized as follows. First, we present the current state of the art by discussing studies that use  \ac{SO} as the main data source and study \ac{ML} topics. Additionally, we discuss some articles about pitfalls and practices of \ac{ML}. Then, in Section three, we present the study design and our approach, followed by the results in Section four, which is divided into subsections aligned with the  categories of practices we found. Each subsection represents an \ac{ML} pipeline stage. Next, we discuss the results, presenting some aspects to bear in mind for some stages that have not been discussed in depth by the \ac{ML}- identified practices. In Section six, we discuss the threats to validity of our study, and finally, we present the conclusion and future work.

%% file: section_related_work.tex
\section{Related Work}
\label{sec:related}

In this section, we discuss the related works, by grouping them in two main categories: (i) studies that use \ac{STE} and focused on \ac{ML}, and (ii) studies that focused on pitfalls, mismatches and practices on \ac{ML}.

\subsection{\ac{ML}-related studies that use \ac{STE} data}

\ac{STE} is a widely used  \ac{CQ+A} website  that contains 173 communities, including StackOverflow \ac{SO}. This website is the ``largest and most trusted online community for developers to learn, share their knowledge, and build their careers''  \citep{stackexchange_2021}. These online communities have been a target of research in software engineering in order to investigate different topics such as (i) trends and impacts in \ac{SO}  \citep[\eg][]{Meldrum_17, Treude_2011, barua_what_2014, Bangash_2019}; (ii) difficulties presented on specific subjects \citep[\eg][]{Alshangiti_2019, Kochhar_2016, Islam_2019} (\eg libraries or software fields); (iii) expertise and reputation of the users in \ac{SO} \citep[\eg][]{Alshangiti_2019, Yang_2014, Vadlamani_2020}; and (iv) APIs and software frameworks (\eg related to issues, problems, documentation) \citep[\eg][]{Islam_2019, Ahasanuzzaman_2020, Han_2020, Hashemi_2020}. ML-related topics have been also investigated with \ac{SO} data \citep[\eg][]{Alshangiti_2019, Islam_2019, Han_2020, Hashemi_2020, Bangash_2019}(see \tabref{tab:state_of_art}). We  discuss some of the most representative works of this last topic in the following paragraphs.

\cite{Alshangiti_2019} used \ac{SO} data to study questions, users' expertise, challenges, and phases in \ac{ML}, by using different research methods. First, they found that the time to answer the questions was ten times longer than typical questions in \ac{SO}, and a higher percentage of ML-related questions did not have accepted answers.  Second, the study compared  \textit{ExpertiseRank}  \citep{Zhang_07} of the users answering the ML-related questions with Web Development respondents and found that respondents for the former are fewer experts. Third, after analyzing the questions in a general way, the authors wanted to understand challenges in \ac{ML}. This analysis was done by studying challenges in a high granularity by examining the distribution of questions in each phase of the \ac{ML} pipeline. They identified that data preprocessing and model deployment were the most challenging phases in the pipeline. In addition, the paper studied challenges in a lower granularity by analyzing them in the questions grouped by topics. These topics were identified using  \ac{LDA} \citep{Blei_2003}, and after analyzing them, ``data'' and ``feature preprocessing'' were identified as the most challenging topics, along with ``neural networks'' and ``deep learning'' as the most popular. Finally, the researchers were interested in identifying which type of knowledge (\ie conceptual knowledge, implementation knowledge, or both types) was needed to address the previously mentioned challenges, discovering that the implementation knowledge type was the most popular in \ac{ML} questions and the most unanswered questions were related to both kinds of knowledge.

Regarding understanding questions, problems, and challenges of \ac{ML} libraries, two studies have investigated these topics.  \cite{Islam_2019} focused their research on understanding questions about popular \ac{ML} libraries (\ie \textit{Caffe} \citep{jia2014caffe},  \textit{H2O} \citep{H2OAutoML20},  \textit{Mahout} \citep{mahout},  \textit{Keras} \citep{Keras},  \textit{MLlib} \citep{ml_lib}, \textit{Scikit-learn} \citep{scikit-learn},  \textit{Tensorflow} \citep{tensorflow2015},  \textit{Theano}  \citep{theano},  \textit{Torch} \citep{torch}, and  \textit{Weka} \citep{weka}). In particular,  their study was related to \ac{ML} libraries that were presented in the \ac{ML} pipeline proposed by \cite{ml_pipeline2017}.  The study considered 3,283 \ac{ML}-related questions of \ac{SO}, which were manually labeled by assigning a particular stage of the \ac{ML} pipeline and then labeling them with subcategories that indicate the post's topic (\ie subjects discussed in posts). Among the most interesting results, they found that the most difficult stage in the \ac{ML} pipeline was model creation for libraries that support clustering, followed by the phase of data preparation. In addition, they found  that (i) type mismatches (\ie the input data does not match the one required by the library) appear in most of the libraries; (ii) shape mismatches (\ie the dimension of the tensor/matrix by a layer does not match the dimension need buy the following layer ) frequently appear in deep learning libraries; and (iii) parameter selection can be difficult in all \ac{ML} libraries.  Finally, the analysis suggests that  some of the libraries had issues in the model creation and training stages. \cite{Hashemi_2020} focused on the documentation of \ac{ML}  software, in which they identified that the questions related to library documentation focused mainly on parameter tuning, model creation, and errors/exceptions. They also found that most of those posts were related to the official TensorFlow  documentation. 




\cite{Han_2020} aimed to analyze what developers talk about three popular \ac{DL} frameworks: 
\textit{Tensorflow}, \textit{Theano},  and \textit{Pytorch}, 
across two popular platforms, \ac{SO} and GitHub. This analysis was based on a set of 26,887 \ac{SO} posts and 36,330 GitHub pull requests and issues that were analyzed with \ac{LDA} to establish their respective topic categories (\eg error, optimization, etc.). These categories were associated with stages of a deep learning workflow, \ac{DL} frameworks, and platforms; the associations allowed analyses at different levels of generalized aggregations, topic, workflow, and framework. The researchers made some comparisons at different aggregations, allowing them to gather insights about \ac{ML} and their respective categories. The authors found that  the most popular (\ie the most discussed stages)  stages were ``model training'' and ``preliminary preparation'' in both platforms and three frameworks. Additionally, the topic that was discussed across all frameworks and platforms was the ``error'' topic. At the framework-platform aggregation, the authors found that the discussion on \ac{SO} posts about workflow phases was similar between \textit{TensorFlow} and \textit{PyTorch}. That  discussion was focused on ``preliminary preparation,'' specifically on the ``error'' topic. On the other hand, the \textit{Theano} posts focused on the ``model training'' stage, especially on the ``error/bug'' topic. When analyzing the GitHub platform, it was found that \textit{TensorFlow} and \textit{PyTorch} issues followed a similar pattern focusing mainly on ``data preparation'' and ``model training'' stages; in contrast, \textit{Theano} issues focused on almost all \ac{DL} stages. Concerning the category-framework level, \textit{TensorFlow} posts centered on ``Optimization,'' \textit{PyTorch} in ``error on data preparation,'' and \textit{Theano} in ``error in model training.'' Moreover,  \cite{Bangash_2019} used \ac{LDA} for extracting possible topics for 28,010 posts and found that the most common topic was ``code errors,'' followed by ``labeling'' and ``algorithms.'' Bangash \etal~   also concluded that most of the theoretical questions were not answered, and the tags that had the original \ac{SO} posts could be improved when using \ac{LDA} topics.

\input{table_state_of_art}

\subsection{Studies on \ac{ML} pitfalls, mismatches and guidelines}
\label{sec:related2}

Previous works have identified common mismatches (\ie problems in the \ac{ML} pipeline due to incorrect assumptions made about system elements)  and pitfalls (\ie a likely mistake or problem) on \ac{ML}.  In particular, these studies aimed to understand and detect mismatches, avoid pitfalls in academic research and identify them in \ac{ML} implementation projects. 

\cite{LewisGrace2021WAIN} studied  mismatches on \ac{ML}-enabled systems. 
They define a \ac{ML} mismatch as ``a problem that occurs in the development, deployment, and operation of an ML-enabled system due to incorrect assumptions made about system elements by different stakeholders: data scientist, software engineer, and operations, that results in a negative consequence'' \citep{LewisGrace2021WAIN}.  The study focused on investigating what common types of mismatches occur in the end-to-end development of ML-enabled systems. To identify and characterize the mismatches, first, they interviewed 20 practitioners to collect mismatch examples, and then the examples were validated with a survey answered by practitioners. After this process, they determined that the majority of mismatches occurred when an incorrect assumption was made about a \textit{trained model}, in particular, assumptions made about \textit{test cases} and \textit{data and API/specifications}.  They also identified, via the survey, that the importance of sharing different information about the trained models varies depending on the role  (\ie data scientist, software engineer, and operations).  Another finding was that the \textit{training data version} (\ie version of the information that is used to train) was the least important category for both interviewees and survey data. They considered this last finding surprising due to the tight relationship between model performance and training data, especially for model retraining and troubleshooting.

\cite{Clemmedsson2018} focuses her research on identifying \ac{ML} pitfalls in four companies with different levels of expertise in \ac{ML} (\ie from almost zero experience to being a company that uses \ac{ML} to define products ). For this, Clemmedsson did a literature review about \ac{ML} stages and \ac{SE}  pitfalls, followed by interviews with participants from the four studied companies to validate the hypotheses generated in the literature review. In the literature review process,  12  pitfalls were identified (\eg overfitting, not testing properly, training data bias, wrong features selection). Clemmedsson also found some similarities with \ac{SE} pitfalls, including \textit{lack of competence},  and \textit{poorly estimated schedules and/or budgets}. Regarding the interviews, some insights were found; for example,  the company with less experience considered that separating a small group of people for an \ac{ML} project is not viable when it is not mainly related to revenue-creating; while the company with greater experience had a challenge regarding being efficient utilizing and drawing conclusions for large amounts of data.  In general, one big pitfall found was not having enough data before starting the implementation of the \ac{ML} project. The four companies also agreed that overfitting and lack of competence (\ie  it is difficult to find/acquire talent with the competence needed for implementing \ac{ML} and data science  projects) are big issues.  In general, they conclude that as experience in \ac{ML} increases,  the overestimation of \ac{ML} capabilities varies. In addition, they concluded that data-related pitfalls are the most severe.

\cite{teschendorff2019avoiding} describes common pitfalls and guidelines when using \ac{ML} in omics (\ie fields of study in biological sciences that ends with -omics, such as genomics) data science. In particular, \cite{teschendorff2019avoiding} highlighted that there is an existing bias in models when the test data leaks into the training and validation sets. They also gave some guidelines to avoid leaking data into the test set, like using different sets for training and testing the models.  

\cite{halilaj2018machine} present common pitfalls and best practices when using \ac{ML} in human movement biomechanics. These issues and guidelines were extracted by four authors from a set of articles that were related to their field. In particular, they described problems and guidelines when splitting a dataset; they also identified that sometimes \ac{ML} is selected to solve problems  in which simpler solutions could be used. Furthermore, they also illustrate that metrics selection is an important part of the design process due to metrics are  used for evaluating a model. 


\cite{biderman2020pitfalls} present a list of pitfalls and recommendations for designing the right algorithm, data collection, and model evaluation, based on their own experience and focused on \ac{ML} research. 
According to the authors, problems may occur because of not engaging the right people in the \ac{ML} development process,  noisy labeling, lack of rigorous evaluation,  and lack of correct comparisons. Regarding the identified practices, they covered topics like integrating stakeholders in the process, collecting data having a hypothesis in mind, and using statistics for quantifying errors in data labeling and in model evaluation. 



\cite{serban2020adoption} list 29  best practices for \ac{ML} applications. These practices were mined from academic and grey literature with an engineering perspective. This means that when mining the practices the authors focused on identifying  practices related to engineering. 
In addition, they grouped the practices into six categories:  \textit{data}, \textit{training}, \textit{coding}, \textit{deployment}, \textit{team}, and \textit{governance}. After identifying the practices, they conducted a survey in which they asked the respondents (\ie researchers and practitioners) about the adoption of the previously identified best practices, and they also surveyed if adopting a set of practices will lead to a desired effect (\ie agility, software quality, team effectiveness, and traceability). As a result they found that larger teams tend to adopt more practices  and \ac{SE} practices tend to have a lower adoption than specialized \ac{ML} practices.



\cite{MichaelLones2021} presents an outline of common mistakes that occur when using \ac{ML} and what can be done to avoid them. These pitfalls and how to overcome them were obtained by Lones while doing \ac{ML} research and supervising students doing \ac{ML} research.  The study is focused on five main stages (i) what to do before model building, (ii) how to reliably build models, (iii) how to robustly evaluate models, (iv) how to compare models fairly, and  (v) how to report results. For each stage, Lones explains what to do and what to do not, always focusing on possible academic results. For example, in  \textit{how to reliably build models}, he explains that test data leakage into the training process has to be avoided. He explains that when this type of leakage happens, data no longer provides a reliable measure of generality and can cause the published models not to generalize other datasets. In this way, Lones presents a study focused on pitfalls based on his experience in academia. 

\cite{wujek2016best} discusse common issues in \ac{ML} and provides guidance to overcome the pitfalls and build effective models. In particular, they focused on data preparation, model training, model deployment, and model monitoring. The described problems related to topics like data leakage, the curse of dimensionality, and overfitting. Besides, the guidelines relate to  how to prepare data (\eg tidy data, standardize, dimension reduction), honest model assessment, regularization, and hyper-parameter tuning, ensemble modeling, aspects to consider when deploying and monitoring a model. 


\cite{horneman2020ai} and \cite{zinkevich_2021} present global (\ie non-detailed) guidelines and practices for \ac{ML} and artificial intelligence (AI). For instance, \cite{horneman2020ai} present 11 AI practices based on their own work, covering topics like expertise in \ac{SE} teams that implement \ac{ML}; design decisions(\eg choosing  a model on needs and not on popularity); security in \ac{ML} systems; recovery and traceability; and ethics. Besides, \cite{zinkevich_2021} gives some guidelines, for engineers, applicable to a general \ac{ML} system and \ac{ML} implementation process.  He divided the guidelines into four main parts, which guide in the understanding of when to build an \ac{ML} system; deploying a first version of an \ac{ML} pipeline; launching and adding new system features, and how to evaluate a model; and refining  the final \ac{ML} models.



\subsection{Summary}

In this section, we presented two main categories of related work. Our first group focused on different aspects of \ac{ML} (\eg difficulties, problems, and frameworks in \ac{ML}), taking as a basis \ac{SO} data. In particular, those studies focused on challenges in \ac{ML} and in its most common libraries, identifying trends and topics, and problems and issues between \ac{ML} platforms, analyzing mainly \ac{SO}. Unlike the first group, this study focuses on analyzing posts from various \ac{STE} pages, not only \ac{SO} posts, and does not focus on frameworks, trends, or topics; our study focuses on analyzing those posts  to identify best practices\footnote{Hereinafter we will refer as best practices for both: good and best practices}  of \ac{ML}. We wanted to explore more than the \ac{SO} community because \ac{ML} practitioners and researchers are not only present in \ac{SO}  but distributed in multiple communities due to the multiple \ac{ML} applications.


In the second group, we presented  a group of studies focused on detecting and avoiding \ac{ML} pitfalls and mismatches based on interviews, surveys, and personal experiences (\ie personal experiences in teaching and research), while we present a list of practices extracted from \ac{STE}, and then were validated by \ac{ML} experts. In particular, \cite{LewisGrace2021WAIN} focused on mismatches in \ac{ML}-enabled systems from three different perspectives (\ie operational, data science, and software engineer), executing an interview followed by a survey to validate their mismatches. Then, \cite{Clemmedsson2018} focused on identifying  \ac{ML} pitfalls, in industry,  based on  a literature review and interviews with four companies.  Next, \cite{teschendorff2019avoiding} and \cite{halilaj2018machine} are studies that focused mainly on pitfalls when using \ac{ML} in their respective fields, when using omic data for data science and  Bio-mechanics, respectively. In addition, \cite{biderman2020pitfalls} presented a series of guidelines based on their own experiences focusing on \ac{ML} research. \cite{serban2020adoption} gave a broader scope of practices for practitioners and researchers, but with an engineering perspective. Then, \cite{MichaelLones2021}  presents an outline of pitfalls and how to avoid them based on their experience of teaching and researching, giving some guidelines. Followed by \cite{wujek2016best}, which present a series of issues and theory about how to do some \ac{ML} tasks in four \ac{ML} phases (\ie data preparation, model training, model deployment, and model monitoring). Finally, \cite{horneman2020ai} and \cite{zinkevich_2021} present non-detailed guidelines for \ac{ML} and AI, based on their own experience. Then, as previously mentioned, we wanted to (i) focus on \ac{ML} practitioners, regardless of their field,  to understand which practices are being discussed in different  \ac{STE}  communities; (ii) our work is more oriented to preventing pitfalls by identifying  best practices and not by identifying at first pitfalls and issues.

%% file: table_state_of_art.tex
\begin{sidewaystable*}
	
	\caption{Previous work using \ac{SO} data focused on \ac{ML}.}
	\label{tab:state_of_art}
	\centering
	\resizebox{\textwidth}{!} {
		\begin{tabular}{llllll}
			\toprule
			\multicolumn{1}{c}{\textbf{Study}} & \multicolumn{1}{c}{\textbf{Year}} &\multicolumn{1}{c}{\textbf{Subtopic}}& \multicolumn{1}{c}{\textbf{ Data}}  & \multicolumn{1}{c}{\textbf{Main Filters}} &  \multicolumn{1}{c}{\textbf{Labeling}} \\
			\midrule
			\cite{Alshangiti_2019} & 2019& Challenges in \ac{ML} & 	 \begin{tabular}[c]{@{}l@{}} \textbf{ Main \ac{SO} data: } \\ \ \ 86,983 questions\\  \ \ 50,630 users\\ \textbf{ Sample \ac{SO} data: } \\ \ \ 684 questions \\ \ \ 50 users \\ \ \end{tabular} &  \begin{tabular}[c]{@{}l@{}} ``Machine learning'' tag, \\ First 25 tags co-ocurred with \ac{ML} tag, \\ Tags exclusively  used  with ML, \\ Posts with active users\\  \end{tabular} & \begin{tabular}[c]{@{}l@{}}\ac{LDA} - topic ID.,\\  Manual \\ \ \ \ac{ML} pipeline stage ID. \\ ExpertiseRank \\ \ \  user expertise\end{tabular}\\
			 
			\cite{Islam_2019}& 2019& \begin{tabular}[c]{@{}l@{}} Challenges  in  \\ \ac{ML} libraries  \end{tabular}  & \begin{tabular}[c]{@{}l@{}} \textbf{ \ac{SO}:} \\ \ \ 3,243  questions  \\ \ \end{tabular}  &  \begin{tabular}[c]{@{}l@{}} Posts with at least one tag: \\ \ \ Tensorflow, Keras, H2o, \\ \ \ Caffe, MLib, Scikit-learn, \\ \ \ Torch, Weka\\  Question's score $\geq 5$ \\ \ \end{tabular} & \begin{tabular}[c]{@{}l@{}} Manual categorizing\\ \ \ 2 classes\\ \ \ 6 sub-classes \\ \ \ 27 sub-sub-classes  \end{tabular} \\

			\cite{Bangash_2019} & 2019& Identifying \ac{ML} topics &\begin{tabular}[c]{@{}l@{}}\textbf{ \ac{SO}:} \\ \ \ 28,010  posts \\ \  \end{tabular} & ``Machine learning'' tag & \begin{tabular}[c]{@{}l@{}} \ac{LDA} - topic ID \\ \ \ K = 50 \\ \   \end{tabular}  \\

			\cite{Hashemi_2020}& 2020 & \begin{tabular}[c]{@{}l@{}} Problems in \ac{ML} \\ libraries documentation  \end{tabular}  & \begin{tabular}[c]{@{}l@{}} \textbf{ \ac{SO}:} \\ \ \ ND sample  \end{tabular}  &  \begin{tabular}[c]{@{}l@{}} Posts with at least one tag: \\ \ \  \ac{ML} \\ \ \ Tensorflow \\ \ \ Pytorch \\ \  \end{tabular}  &  \begin{tabular}[c]{@{}l@{}} Manual\\ \ \ Problem categorizing \\ K-means \\ \ \ ID expertise groups \\ \ \end{tabular} \\

			 \cite{Han_2020} & 2020& \begin{tabular}[c]{@{}l@{}} Issues in  \\ \ac{ML} libraries\\ between platforms \end{tabular}   & \begin{tabular}[c]{@{}l@{}} \textbf{ \ac{SO}: data} \\ \  \ 23,908 question \\ \textbf{ GitHub: data}  \\ \ \ 7,987 pull requests \\ \ \  12,400 issues\end{tabular}   & \begin{tabular}[c]{@{}l@{}} \textbf{\ac{SO}} \\ Posts with at least one tag: \\ \ \ tensorflow \\ \ \ pytorch \\ \ \ theano \\ \textbf{GitHub} \\ Data from main repositories: \\ \ \ tensorflow \\ \ \ pytorch \\ \ \ theano\end{tabular}&  \begin{tabular}[c]{@{}l@{}} \ac{LDA} - topic ID \\ \ \ k = 75 \\ Manual\\ \ \  match each \ac{LDA} \\ \ \ topic  to a \ac{DL} \\ \ \ pipeline stage \end{tabular} \\
			\bottomrule
			\multicolumn{4}{l}{ID is used as an acronym od identification. ND is used as an acronym of  not defined }
	
		\end{tabular}
		}
\end{sidewaystable*}




%% file: section_design_study.tex
\section[final]{Study Design}
\label{sec:design}
This study aims to analyze which best practices of \ac{ML} are discussed on  14  \acp{STE} websites. The purpose is to define a taxonomy of \ac{ML} best practices discussed on the  \acp{CQ+A} websites and highlight the corresponding \ac{ML} pipeline stages in which they are located. Note that this study is conducted from the perspective of researchers interested in building a catalog of \ac{ML} best practices that are used by practitioners. This catalog could also be useful for researchers from different disciplines, including software engineering, that are interested in using \ac{ML} as part of their proposed approaches. The context of this study  consists of 157 best practices extracted from  121 Posts (\ie pairs of questions and answers (242 posts in total))  from 14 different  \ac{STE} websites. These practices  were validated by four \ac{ML} experts.  All the data used in this study are available in our \textit{online appendix} (\emph{\textbf{\url{https://tinyurl.com/bestPML}}}). This study investigates the following research question: 

\begin{center}
\textbf{RQ: \textsl{What are the \ac{ML} best practices found in \ac{STE} websites?}}
\end{center}

\begin{figure}
	\centering
	\includegraphics[width=0.9\linewidth]{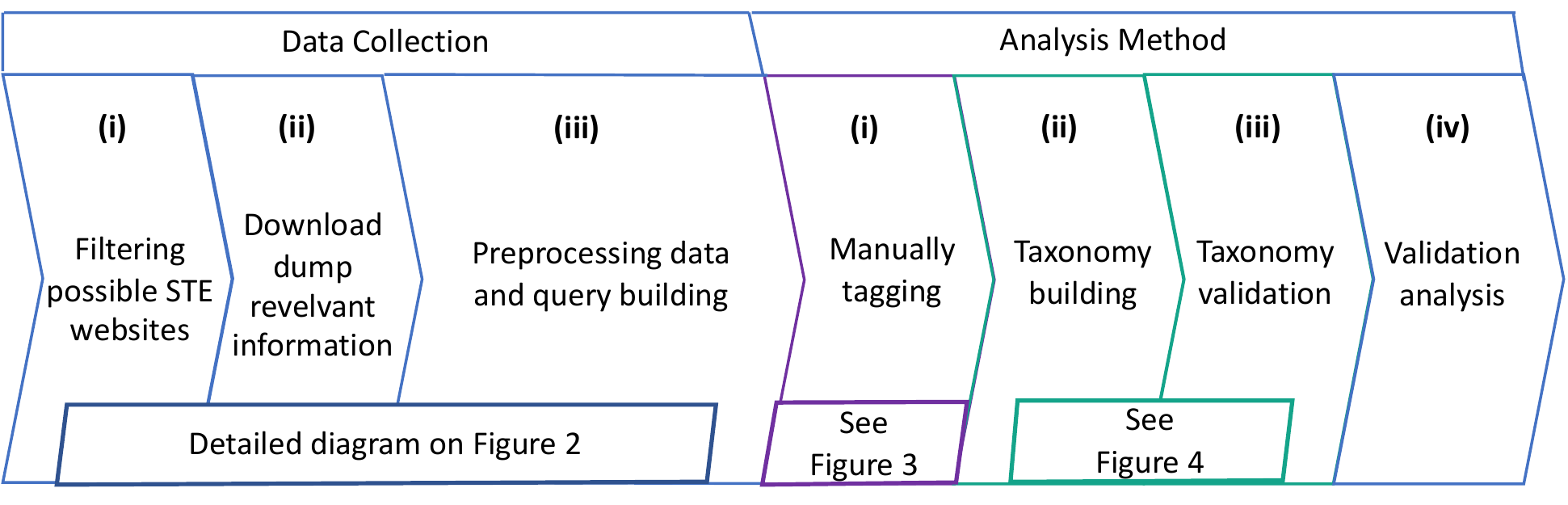}
	
	\caption{Overview of the methodology.
	}
	\label{fig:method}
	
\end{figure}

To answer the research question, we executed the following steps (see \figref{fig:method}): 

\begin{itemize}
	\item Data Collection:
	\begin{itemize}
	\item [i.] Select some \acp{CQ+A} \textit{website} as target pages to extract information.
	\item [ii.] For each selected \textit{website}, we download their dump(s) and preprocess data.
	\item [iii.] Build queries to extract relevant \textit{posts}.
    \end{itemize}
	\item Analysis Method:
	\begin{itemize}
	\item [i.] Manual tagging of the resulting \textit{posts}, and identification of   possible best practices and the corresponding stage in an \ac{ML} pipeline.
	\item [ii.] Build a taxonomy of possible best practices.
	\item [iii.]Validate the taxonomy with four \ac{ML} experts.
	\item [iv.]Analyze the decision of each expert about each practice, allowing us to establish a set of best practices discussed in different \ac{STE} communities.
	\end{itemize}
\end{itemize}

We describe each step in detail in the rest of this section.

\vspace{-10pt}

 \subsection{Data Collection}

\begin{figure}[h]
 	\centering
 	\includegraphics[width=0.9\linewidth]{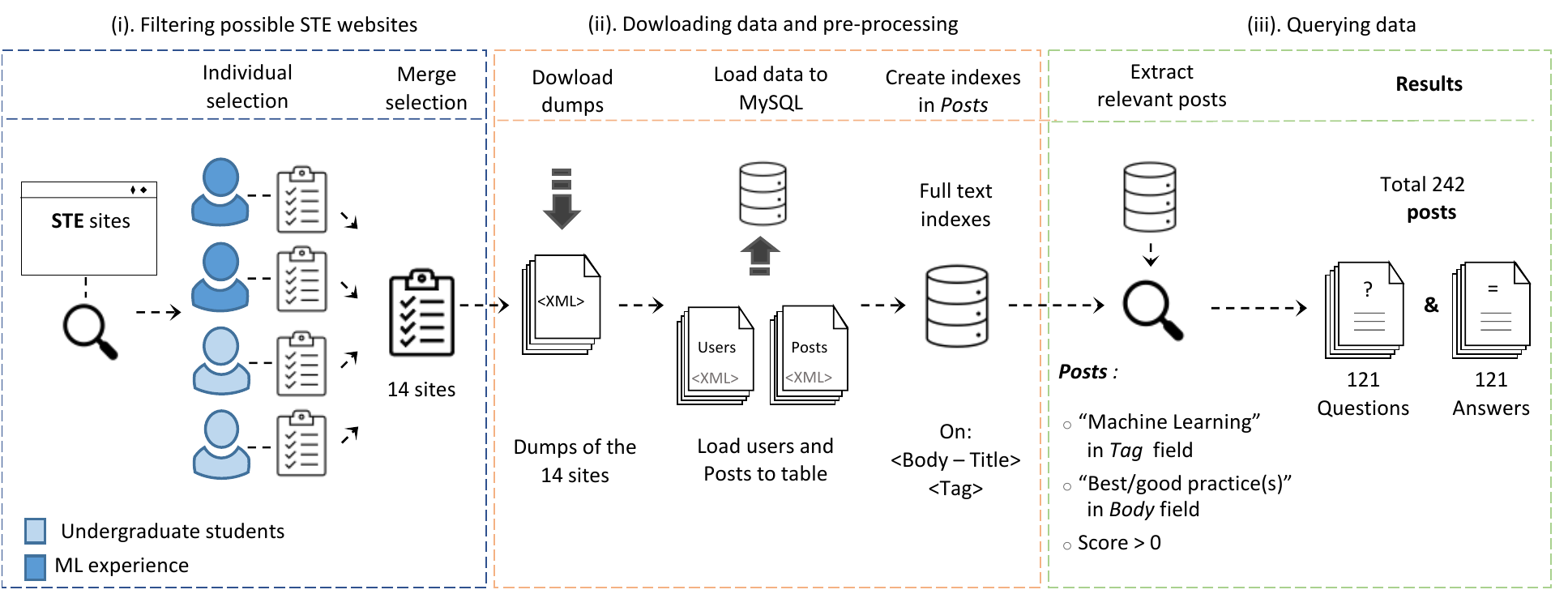}
 	
 	\caption{Data collection procedure.
 	}
 	\label{fig:data_collection}
 	
 \end{figure}

In this phase, we collected and filtered relevant posts of \ac{STE} websites (see \figref{fig:data_collection}). Four people (3 authors and an external collaborator) identified a list of pages from \ac{STE} websites\footnote{https://stackexchange.com/sites} that could could contain posts related to best practices in \ac{ML} (see \tabref{tab:experience}). Each person listed the pages they considered relevant, then the four people reached an agreement leaving a total of 14 sites (see \tabref{tab:websites_ml}).  Once the sites were selected, we proceeded to download the respective dumps from the \ac{STE} data archive\footnote{https://archive.org/details/stackexchange\textunderscore 20210301}\!. Each dump has multiple \textit{XML} files related to relevant information (\eg posts, users), except \ac{SO}, which has a separate dump for the posts' information. In this particular case (\ie \ac{SO}), we only downloaded posts' dumps.  The relevant information (\ie posts.xml) was uploaded into tables in a local MySql database.

 
 \input{table_wesites_ml} 

 After all the data was stored in the database, we began building queries to extract only those posts that were interesting for our study. To retrieve only relevant information, we built a query that filtered   \texttt{posts} that met the following requirements:
 \newpage
 \begin{enumerate}[label=\roman*.]
 	\item Posts that are indeed questions or answers  (\eg we removed those who describe a tag).
 	\item Posts with a score equal or greater than one.
 	\item Questions that have  a ``Machine Learning'' tag.
 	\item Questions that have in their body or title  ``best practice(s)'' or ``good practice(s)''.
 \end{enumerate}

For each question that met those requirements, we retrieved the accepted and the most voted answer with a score greater than one. We retrieved both answers because, in some cases, the most voted answer  is not the same as the answer accepted by the user who published the question.  \textit{After executing the queries, we obtained 121 pairs of questions and answers, which means a total of 242 posts}. \tabref{tab:websites_ml}  depicts how many posts were retrieved for each \ac{STE} selected page. More details about table building and query creation can be found in the online appendix. 

\input{table_experience_people}

 \subsection{Analysis Method}
 \vspace{-5pt}
In order to answer the research question, we followed an open-coding inspired procedure \citep{corbin1990grounded} to extract the best ML practices reported by practitioners on the \ac{STE} pages. We divided this process into four main phases:  (i) manual tagging; (ii) taxonomy building; (iii) taxonomy validation; and  (iv) result analysis.

\subsubsection{Manual tagging}
   
\begin{figure}
	\centering
	\includegraphics[width=0.9\linewidth]{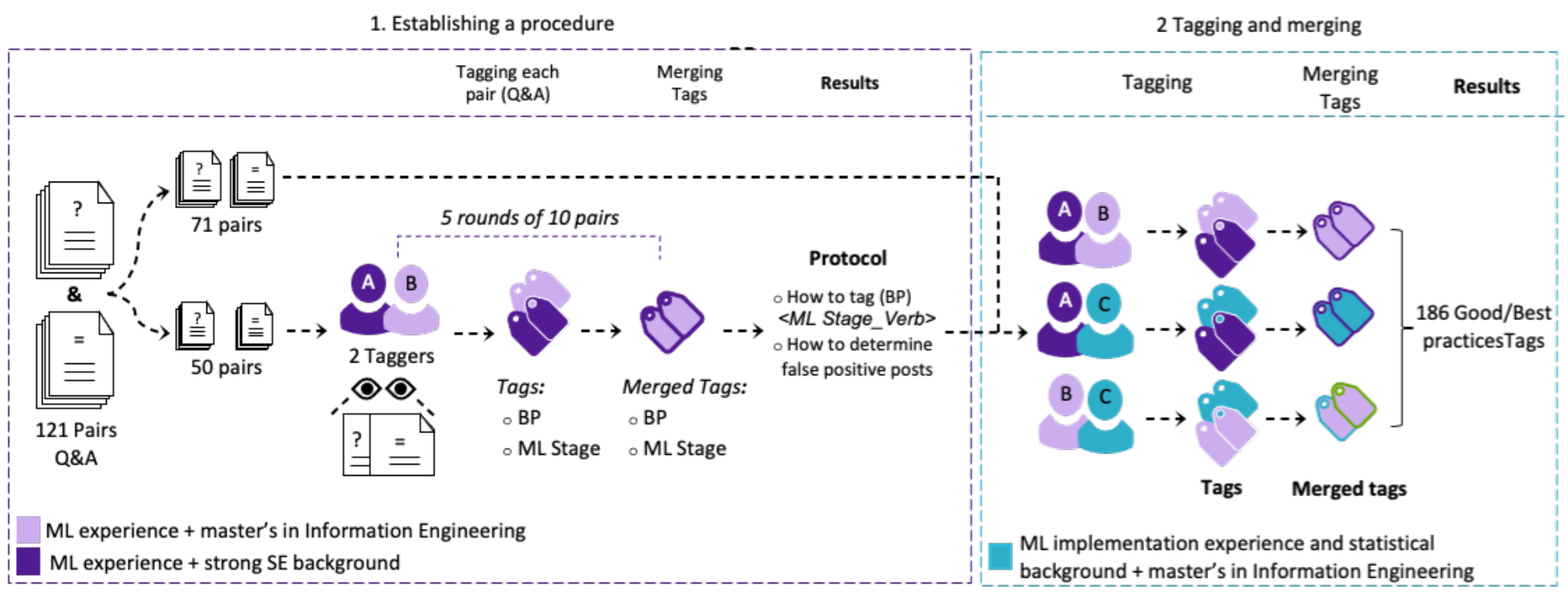}
	
	\caption{Analysis method process Part (i)
	}
	\label{fig:analysis_method}
	\vspace{-10pt}
\end{figure}

    \vspace{-5pt}
At this stage, the 121 pairs of posts were manually tagged in order to extract for each pair of posts (i) the related \ac{ML} pipeline stages (\cite{amershi2019software}); (ii) the external references; and (iii) the identified best practices.  This stage was divided into three sub-phases: establishing a procedure, tagging, and merging (see \figref{fig:analysis_method}).

\textbf{Establishing a procedure.} We defined (i) how to assign an \ac{ML} stage from the ones defined by \cite{amershi2019software},  (ii) how to extract external references included in each pair of posts; and (iii) how we were going to tag the  ML practices reported (\ie how we were going to represent each practice with a tag) in each pair of  question-answer. In this phase, two authors (see \tabref{tab:experience}) and an initial set of 50 pairs  (\ie 50 pairs out of 121) of question-answer were involved.  The tagging process was supported by a self-made web application. This application  showed for each pair of question-answer  a series of fields to fill out the aforementioned information and also a question that allowed to identify whether the pair was a false positive (the pair did not contain any best \ac{ML} practice).

This phase was executed in five rounds of tagging and merging. We did this process in batches in order to make easier the merging process and  also to merge possible similar tags as soon as possible.  In the tagging process, each author identified for each pair of question-answer:
  
   \begin{enumerate}[label=\roman*.]
  	\item Machine learning pipeline stages  (\ie requirement definition, data collection, data labeling, data cleaning, feature engineering, model training, model evaluation, model deployment, model monitoring) related to the posts
  	\item External references to each post, whether they were grey or white literature.
  	\item Possible best practices related to each post, by using tags. Each tag had the following structure $<$\ac{ML} Phase category$>+ \rule{2ex}{.4pt}+<$Verb$>+\rule{2ex}{.4pt}<$ complement$>$, in which $<$Verb$>$ indicates an action that could happen in the category. One example of a tag with that format is ``\textit{Training\_use\_dropout\_layers}''. Note that \textit{Phase Category}, could be an exact stage of the \ac{ML} pipeline or relates to it.
  \end{enumerate}
  

In each of the merging steps of this phase (\ie one merging step per round, 5 in total), each author exposed their tags and data for the three aforementioned elements (\ie stage, references, and practices)  and then solved conflicts.  The merging process involved analyzing how different was the information provided by each tagger. On the one hand, if both taggers had the same tag(s) and stage(s), they continued with the following pair of question-answer. On the other hand, three main scenarios could happen: (i) the tags were similar (\ie this means that the tags were related to the same practice but with different words); (ii)  the tags were complementary (\ie it means that each tagger identified different tags, and both tags were valid); (iii) the point of view of both taggers were opposite (\eg the first tagger considered that the post was not a useful post for the investigation, but the second tagger did consider it a useful post). In the first two cases, the taggers agreed to a mutual tag(s) and stage(s). In the last case, each tagger exposed their point of view and then  decided to join tags and stages. 
  
At the end of this phase, we identified that (i) pairs of question-answer could be related to more than one \ac{ML} pipeline stage; (ii) the external references could reference other posts, articles, or books;  and (ii) there was a question-answer pair about a best practice in a specific language (\eg Python, Java, Julia), thus, the post was labeled as a false positive.

\textbf{Tagging and Merging.} 
We executed two rounds of tagging and merging ensuring that all of the 71 remaining pairs  were tagged by at least  two people of a group of three possible taggers (see \tabref{tab:experience}).  

In the first round, 60 pair of post were analyzed.  Twenty  posts were assigned to taggers $Cl_A$ and  $Cl_B$, another 20 to taggers $Cl_A$ and $Cl_C$, and the other 20 to taggers $Cl_B$ and $Cl_C$.  We followed the same procedure for the second round to label the remaining 11 pairs~\footnote{We did not divide evenly due to (i) the number of tags were even, and (ii) one of the taggers had time limitations}. Remember  that in the tagging process, we do not only identify the best practices but also the related  \ac{ML} pipeline stages, and the external references mentioned  in the post. In the tagging process, when the taggers identified a possible best practice mentioned in a question or answer, we did it without judging the practice, \ie the taggers did not discuss whether it was a real best practice in \ac{ML}. This validation process was later done by four ML experts. 

After each taggers group ended a tagging round, the involved taggers met and solved conflicts  by  homogenizing  similar practices (\eg the group conformed by the taggers $Cl_A$ and $Cl_B$ ended, then the tagger $Cl_A$ met with $Cl_B$ to solve conflicts).  Following the aforementioned procedure for solving conflicts in the previous phase.  

\emph{{At the end of this phase, we obtained 186 different tags describing \ac{ML} best practices discussed by practitioners that are reported as part of questions or answers in the \ac{STE} posts.  We also identified that 35 pairs of question-answer  were false positives.}}
  
  \vspace{-5pt}
\subsubsection{Taxonomy building }
  \vspace{-5pt}

\begin{figure}[h!]
	\centering
	\includegraphics[width=0.9\linewidth]{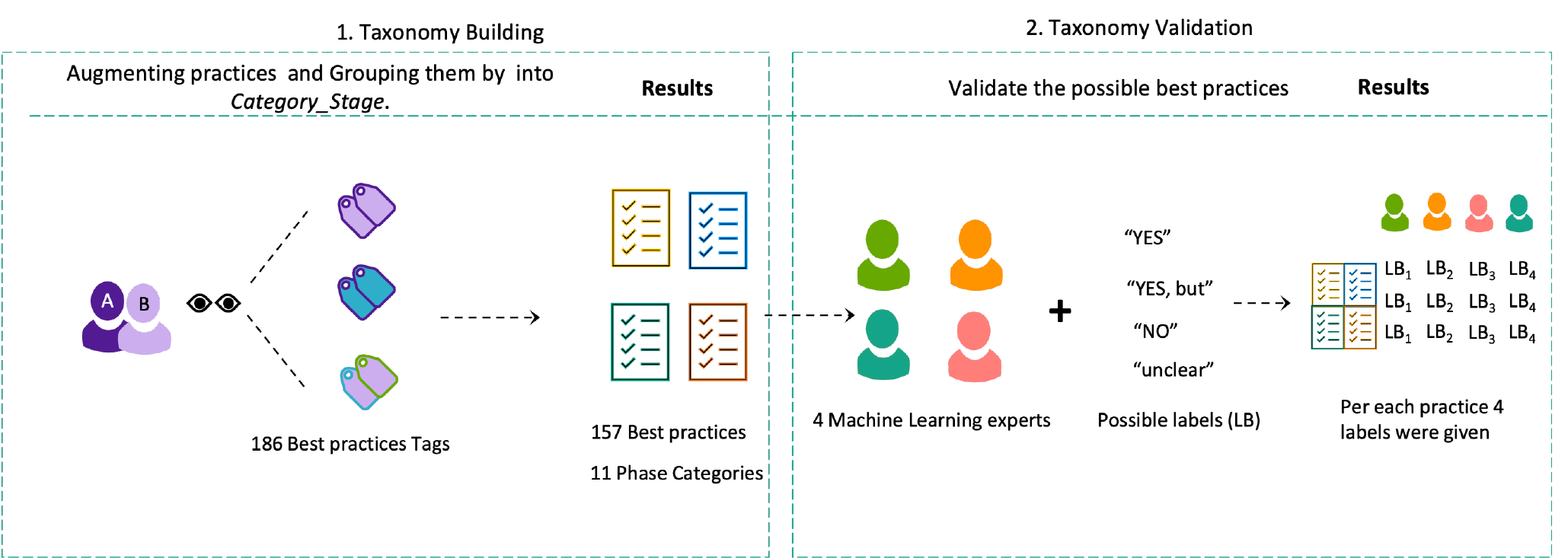}
	
	\caption{Analysis method process Part (ii) and Part (iii)
	}
	\label{fig:analysis_method2}
	
\end{figure}

 Two authors ($Cl_A$ and $Cl_B$ from the previous phase) built a taxonomy by grouping the 186  previously obtained  tags. The tags that represented the best practices were augmented with a verbose description, based on the text in the original post (see \figref{fig:analysis_method2}). After having the 186 different tags, we grouped the tags by phase and verb established in the tag (remember that each tag followed this structure $<$\ac{ML} Phase category$>+\rule{2ex}{.4pt}+<$Verb$>+\rule{2ex}{.4pt}+<$ complement$>$). The phase category could be an exact stage of the \ac{ML} pipeline or relates to it.  At this point, some phase categories were merged if they refer to the same subject but with different words (\eg wrangling and pre-processing). A similar action was taken with similar verbs (\eg replace missing values and impute missing values).

Then, we converted each tag, that represents a best practice, into a verbose text explaining the identified practice. In this process, we noticed that some tags  were ``main tags'' (\ie tags that explained a general practice that could be instantiated/exemplified) and others were instances/examples of that main practice. In addition, we noticed that some existing tags related to the same best practice but were not detected in the merging process.

\textit{At the end of this phase, a taxonomy with 3 levels (\ie phase category\footnote{In some particular cases, due to readability aspects the verb was changed to a complete sentence}, verb, and practice)  were defined and homogenized; we obtained  11 phase categories, 25 verbs and 157 descriptions of potential best practices. }

\vspace{-5pt}

\subsubsection{Taxonomy Validation }

In this phase (see \figref{fig:analysis_method2}), four \ac{ML} experts (see \tabref{tab:experience}) validated the taxonomy  (\ie the 157 possible best practices) by giving their expert opinion on which of the practices reported in the taxonomy should be considered as best practices. 

In the execution of this phase, each expert had the possibility of tagging each practice with four labels ``YES,'' ``YES, but'', ``NO'',  and ``Unclear''. The first label indicates that the practice was a true best one.  The second label, ``YES, but'', indicated that the practice was a good one, but some clarifications or comments should be considered. Hereinafter, we will refer to the practices that receive either of the previous two labels as validated. The third label, ``NO'', indicated that the practice was not a best practice; the last label,``unclear'',  was added for the cases in which the practice description was not clear for the expert. \textit{At the end of this phase, each of the four experts gave a label to each practice and clarification if it applies.}

\vspace{-5pt}
\subsubsection{Validation analysis }

After the previous stage, performed by the  four \ac{ML} experts, we selected the best practices validated by them. In particular, we selected those practices in which the majority of the experts, \ie at least three experts said ``YES'' or ``YES, but''. As a result, from the 157 original practices, 127 were selected as validated best practices (81.5\%). \textit{From those 127 practices, 58 were validated as best practices by three of the experts, and the remaining  70 by the four experts. In addition, if a comment was given by an expert, we analyzed  it and discussed the comment in the results section.}

%% file: table_wesites_ml.tex
\begin{table}[b]
	\vspace{-5pt}
	\centering
	\vspace{-5pt}
	\caption{Selected \ac{CQ+A} websites.}
	\vspace{-5pt}
	\label{tab:websites_ml}
		\begin{tabular}{lcc}
			\toprule
			\multicolumn{1}{c}{\textbf{\ac{CQ+A} Sites} } & \multicolumn{1}{c}{\textbf{Total Q\&A}} &
			\multicolumn{1}{c}{\textbf{Total selected Q\&A}} \\
			\midrule
			
			Stack Overflow &  52.059,483 & 114 \\
			Cross Validated* &  346,009 & 64 \\
			Data Science &  58,718 & 46 \\
			Code review &  179,877 & 6 \\
			Artificial Intelligence &  16,838 & 4 \\
			Computer Science &  85,202 & 4 \\
			Open Data &  12,228 & 2 \\
			Software Engineering  &  226,502 & 2 \\
			Internet of Things &  4,090 & 0 \\
			Theoretical Computer Science &  26,672 & 0 \\
			Engineering &  410,450 & 0 \\
			Electrical Engineering &  25,537 & 0 \\
			Computational Science &  21,427 & 0 \\
			Signal Processing &  49,604 & 0 \\
				
			\bottomrule
			\multicolumn{3}{l}{*The official name in the \ac{STE} is Statistical Analysis}
		\end{tabular}
\end{table}


%% file: table_experience_people.tex
\begin{small}
\begin{table}[H]

	\centering
	\caption{\begin{small}Roles and experience of people involved in the study.\end{small}}
	\label{tab:experience}
	\setlength{\tabcolsep}{3pt}
		\resizebox{\textwidth}{!} {
	\begin{tabular}{m{25pt} m{145pt} m{20pt} m{20pt} m{20pt} m{20pt}} 
		\toprule
		\multicolumn{1}{M{27pt}}{\textbf{Person}} & \multicolumn{1}{c}{\begin{tabular}[c]{c} \textbf{Experience} \end{tabular}}& \multicolumn{1}{M{39pt}}{ \textbf{\ac{CQ+A}}   \textbf{Selection}}  & 	
		\multicolumn{1}{m{35pt}}{\textbf{Manual}   \textbf{Tagging}} &\multicolumn{1}{M{40pt}}{ \textbf{Taxonomy}  \textbf{building}}&\multicolumn{1}{M{50pt}}{\textbf{Taxonomy}   \textbf{Validation}}\\
		\midrule
		
		$U_1$&  Previously taken Business Intelligence undergraduate course.   Masters' \ac{ML} formal course &\checkmark &\xmark&\xmark&\xmark\\ \cmidrule[0.05em]{2-6}
		$U_2$&  Previously taken Business Intelligence undergraduate course.   Masters' \ac{ML} applied course with emphasis in ethics&\checkmark&\xmark&\xmark&\xmark\\
		\cmidrule[0.05em]{1-6}
		$Cl_A$& \ac{ML} implementation in \ac{SE} projects and  master's in  information Engineering&\checkmark&\checkmark&\checkmark& \xmark\\ 	\cmidrule[0.05em]{2-6}
		$Cl_B$&  \ac{ML} implementation in research projects and a strong \ac{SE}  background &\checkmark&\checkmark&\checkmark& \xmark\\ 	\cmidrule[0.05em]{2-6}
		$Cl_C$&\ac{ML} implementation, a statistical undergraduate background and a master's in Information Engineering. &\xmark&\checkmark&\xmark&\xmark\\ 	\cmidrule[0.05em]{1-6}
		$Exp_1$&\ac{ML} research and \ac{ML} projects (22 years), Interdisciplinary projects Information systems for decision making in bioanalysis, biology, land management and agriculture, health  &\xmark&\xmark&\xmark&\checkmark\\ \cmidrule[0.05em]{2-6}
		$Exp_2$& \ac{ML} research and \ac{ML} projects, industry related experience (25 years).  Leads the \ac{ML} perception and Discovery lab (MindLab)&\xmark&\xmark&\xmark&\checkmark\\ \cmidrule[0.05em]{2-6}
		$Exp_3$&  In academia (three years) and  in industry (four years) in \ac{ML} related&\xmark&\xmark&\xmark&\checkmark\\ \cmidrule[0.05em]{2-6}
		$Exp_4$& In academia and industry a total of six and a half years in \ac{ML} related topics &\xmark&\xmark&\xmark&\checkmark\\
		\bottomrule \addlinespace[2pt]
		\multicolumn{6}{l}{$U_i$: Undergraduate student, $Cl_i$: Collaborator, $Exp_i$: \ac{ML} expert }

	\end{tabular}
}
\vspace{-5pt}
\end{table}

\end{small}

%% file: section_result.tex

\section{Results: what are the \ac{ML} best practices found in \ac{STE} websites?}
\label{sec:results}

In this section, we present the results achieved by  our open-coding procedure, described in the previous section, and the subsequent validation by \ac{ML} experts. We present the 127 validated \ac{ML} best practices, \ie the practices reported by practitioners on \ac{STE} websites and selected as best practices by the majority of the four experts involved in our study (\ie at least three experts). For each practice, we provide examples, extended explanations, and extra considerations if applicable. In addition, we report references to existing literature that could be used by the readers to get more knowledge about how and when  to apply the practices.

\begin{figure}[h]
	\centering
	\includegraphics[width=\linewidth]{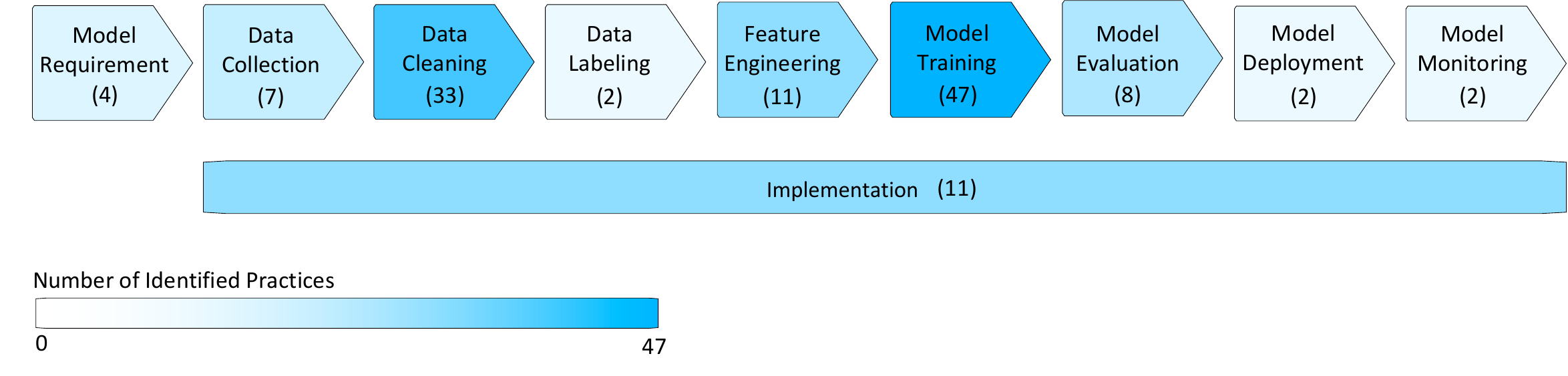}
	
	\caption{Number of practices per \ac{ML} pipeline stages. Image bassed on the pipeline  figure from  \citep{amershi2019software}, a heat map that represent the number of identified practices per stage was added.  Notice that this is not a linear pipeline, there are feedback loops between  stages.} 
	\label{fig:stages}
\end{figure}

Before going into the practices' details, we present an overview of the obtained results. \figref{fig:stages} depicts the number of practices identified per ML pipeline stage \citep{amershi2019software}. Moreover, \tabref{tab:summary} summarizes the covered topics by the validated practices. With this, we do not mean that all the topics present  in the table are fully embraced, but at least one practice is related to them. Note that we added an additional stage, entitled Implementation, in addition to the ones provided by Amershi \etal \citep{amershi2019software}. 

In the rest of this section, we present detailed explanations and examples of each  \ac{ML} validated practice per stage.  When describing the practices, we refer to each one by using an acronym composed of the \emph{stage} id and a number  $i$ ($stage_i$). For instance,  ($mr_1$) means that it is the 1$^{st}$ practice in the model requirements (\emph{mr}) stage. In each subsection, we present first all the practices validated by four experts and grouped by topics; next, we present the practices validated by three experts. For each practice, we present a table with the id, short description, number of experts that validated the practice, and extra considerations. Those extra considerations have two origins, experts' comments  given when  validating the practice or clarifications  that we added.  In the case of Data Cleaning and Model Training, we divided  the stages into subcategories to make the reading easier because, in those cases, there is a large number of validated practices.

\input{table_summary_topics}

\subsection{Model Requirements (mr)}
\label{subsec:mr}

In this stage, designers decide  the functionalities that should be included in an ML system, their usefulness for new or existing products,  and  the most appropriate type of \ac{ML} model for the expected system features \citep{amershi2019software}.  Four \ac{ML} best practices were identified for this stage (see \tabref{tab:pr_mr}). 
\newpage

\input{table_practices_mr}

The first practice (\acron{mr}{1}) relates to metrics selection, and it states that  \practice{the goal of the model that is going to be trained should be considered when selecting  the most appropriate metrics for assessing the model performance/accuracy}. 
For instance, if the goal/requirement is to build a more precise model  than one that retrieves all the relevant instances (\ie all the instances that are important for the task), then a metric that focuses on precision is needed, like \textit{precision} or \textit{f-score} with a $ \beta $ coefficient that favors precision. For example, for the task of classifying X-rays as images with malignant tumors,  some of the decisions to take into account would be (i)  would it be better to retrieve with the model all the  images with malignant tumors (\ie relevant images), regardless some of the retrieval images would be  false positive (\ie images classified as relevant ones, but in reality, they do not have malignant tumors)?, or (ii)  would it be better to retrieve some of the relevant images with high certainty, decreasing the false-positive images, but also missing to retrieve some images  with malignant tumors (\ie relevant images)?; or (iii) is it desirable to have a model that balances the  two aforementioned cases?.  In the first case, the model would be focused on \textit{recall} (\ie retrieving  as many relevant images as possible); in the second case, the classification model would be focused on \textit{precision}; and in the third case, the model would be focused on balancing the two  aforementioned metrics (\ie \textit{f-score}). 

Aligned with the \acron{mr}{1} practice, the (\acron{mr}{2}) practice suggests that \practice{it is important to identify the needs/requirements of model retraining, \eg whether it should be online, periodically, or in batch}. For instance, if you need to update a model each time that a user interacts with the model or when the data is being updated, an online approach is suitable for this case.  The third practice (\acron{mr}{3}) relates to a model execution:  \practice{if a ML model  is published as a cloud service,  specifically, when a classification/prediction task uses ML cloud-based services, it is important to define the use case and the model requirements in order to identify how frequently the service should be invoked}. 
This is in particular important for selecting the model and for provisioning the infrastructure required to execute the model. For instance, if a cloud model is expected to constantly classify posts from social networks, then it requires a more performant model and infrastructure than a use case in which the  model  classifies a group of posts at the end of each day.

The fourth practice (\acron{mr}{4})  indicates that \practice{when using \emph{probabilistic forecasting} in a decision system, it is necessary to decouple the probabilistic model optimization from the probability threshold selection}. 
In other words, the model hyper-parameters should be trained first, and then the decision threshold should be adjusted. Adjusting this threshold will trade-off  between \textit{True Positives} (\textit{TP}) (\ie  number of positive instances correctly classified) and \textit{False Negatives} (\textit{FN}) (\ie  number of negative instances incorrectly classified), as well as  \textit{False Positives} (\textit{FP}) (\ie number of positive instances incorrectly classified)  and \textit{True Negatives} (\textit{TN}) (\ie number of negative instances correctly classified).  In other words, adjusting the threshold could cause that some of the instances classified as TP could be classified as FN and the same with FP and TN. To illustrate this, imagine that a logistic regression model is used for identifying vulnerable or not vulnerable code snippets.  As \acron{mr}{4} indicates, in first instance the logistic regression should be trained in order to find the optimum hyper-parameters (\eg solver, penalty, and $C$). Then,  the decision threshold should be adjusted. Now, let's assume that we have a  binary decision system (\eg fix or not a potentially vulnerable code) and  that the vulnerabilities that are being detected have a low score in the  \textit{CVSS} \footnote{The common vulnerability Scoring system (https://nvd.nist.gov/vuln-metrics/cvss)} rating.  Then, if some vulnerable code snippets are being classified as not vulnerable (FN), thus, those snippets will not be fixed. That will be less crucial  than if those vulnerabilities had a critical score in the same ranking. In consequence, in this case, adjusting the threshold in order to trade-off between TP for FN and FP for TN  
will be useful depending on the cost of classifying  an instance incorrectly.  On the other hand, we could also be in a scenario in which we have a binary target (\eg vulnerable or not) but multiple decision thresholds associated with it  (\eg not fixing it, fixing it in one hour, one day, or in a week). In this second case, each threshold, associated with a fixing time period has to be also adjusted after optimizing the  logistic regression.  Note that three authors validated the \acron{mr}{4} practice; however, one expert answered with ``YES, but''. The clarification given by the experts is that ``I think this GP belongs to the training category''.  This practice is a borderline between both stages (Model Requirement and Training). However, since the other three experts commented anything about the practice's location, we decided to keep  the practice at this stage. 

\subsection{Data Collection (dc)}

This second stage encompasses  looking for, collecting, and integrating available datasets \citep{amershi2019software}. Datasets can be created from scratch, or  existing datasets  can be used to train models in a transfer learning fashion. Both scenarios are widely used when creating ML systems.  In this stage, seven validated practices were identified (see \tabref{tab:pr_dc}). 
Bear in mind that the identified practices relate to some characteristics that the collected data has to meet during/after this process and not to the collection process itself.

\input{table_practices_dc}

%% file: table_summary_topics.tex
\begin{small}
	\begin{table}
		\centering
		\caption{\textbf{Summary}. Topics addressed in the identified practices per stage.}
		\label{tab:summary}
		\setlength{\tabcolsep}{2pt}
			\begin{tabular}{cc}
				\toprule
				\multicolumn{1}{c}{\textbf{Stage}} & \multicolumn{1}{c}{\begin{tabular}[c]{@{}c@{}}\textbf{Topics Covered} \end{tabular}}\\
				\midrule
				\multicolumn{1}{c}{ \textbf{Model Requirement (mr)} } &\multicolumn{1}{l}{\begin{tabular}[c]{@{}l@{}}
						\tabitem Identification of model retraining needs and goals \\
						\tabitem Frequency call of external services \\
						\tabitem Metric selection \\
						\tabitem Decision involving Probabilistic models\\
				   \end{tabular}}\\  \cmidrule[0.05em]{1-2}
			   
				\multicolumn{1}{c}{ \textbf{Data Collection (dc)} } & \multicolumn{1}{l}{\begin{tabular}[c]{@{}l@{}}
					\tabitem Data distribution \\
					\tabitem Composition of images\\
					\tabitem Measuring size of datasets\\
					\tabitem Augmenting data \\
					\tabitem Characteristics of positive and negative  samples\\
					\tabitem Object detection considerations
				\end{tabular}}\\ \cmidrule[0.05em]{1-2}
			
				 \multicolumn{1}{c}{ \textbf{Data Cleaning (dcl)} }& \multicolumn{1}{l}{\begin{tabular}[c]{@{}l@{}} 	
						\emph{Exploratory data analysis (subcategory)} \\
					    \tabitem Determining types of variables and dependency between them\\
						\tabitem Understanding geographical data\\
						\tabitem Identification trends and errors  in data\\
						\emph{Data (subcategory)}\\
						\tabitem Splitting dataset  for specialized models \\
						\tabitem How to split a data set  for the next stages\\
						\tabitem When to measure size of a dataset \\
						\emph{Wrangling (subcategory)}\\
						\tabitem Encoding variables  (numerical and categorical)\\
						\tabitem Imputation techniques\\
						\tabitem Transformation  (\eg Normalization, Scaling) \\
						\tabitem Anonymization
				\end{tabular}}\\ \cmidrule[0.05em]{1-2}
			
			 \multicolumn{1}{c}{ \textbf{Data Labeling (dl)}} &\multicolumn{1}{l}{\begin{tabular}[c]{@{}l@{}} 	
						\tabitem Scalability for labeling data\\
						\tabitem  Parameterize region of interest in object detection\\
				\end{tabular}}\\  \cmidrule[0.05em]{1-2}
			
			\multicolumn{1}{c}{  \textbf{Feature Engineering (fe)} } & \multicolumn{1}{l}{\begin{tabular}[c]{@{}l@{}} 	
						\tabitem Dimensional reduction \\
						\tabitem  Feature selection \\
						\tabitem  Feature selection assumptions\\
				\end{tabular}}\\  \cmidrule[0.05em]{1-2}

				\multicolumn{1}{c}{ \textbf{Model Training (mt)}} &
				\multicolumn{1}{l}{\begin{tabular}[c]{@{}l@{}} 	
						\emph{Learning phase (subcategory) }\\
							\tabitem  Training specialized models \\
							\tabitem Invariant-algorithms when different scale systems \\
							\tabitem Avoid overfitting \\
							\tabitem Neural network initialization  and convergence\\
							\tabitem Faster convergence in Multi layer perceptrons \\ 
						
						\emph{Validation phase (subcategory)}\\
						\tabitem  Optimization hyper-parameters \\
						\tabitem Usage of different types of data splits for hyper-parameter tuning \\
						\tabitem Prevent overfitting\\
				\end{tabular}}
				\\  \cmidrule[0.05em]{1-2}

				\multicolumn{1}{c}{  \textbf{Model Evaluation (me)} }&
				\multicolumn{1}{l}{\begin{tabular}[c]{@{}l@{}} 	
						\tabitem  Cross-validation for evaluation \\
						\tabitem  Evaluating bias using a superset vocabulary \\
						\tabitem  Unit testing \\
						\tabitem  Robustness \\
				\end{tabular}}
				\\  \cmidrule[0.05em]{1-2}
				
				\multicolumn{1}{c}{ \textbf{Model Deployment (md)} }&
				\multicolumn{1}{l}{\begin{tabular}[c]{@{}l@{}} 	
				\tabitem  Retraining a model for deployment \\
				\tabitem  Exporting the \ac{ML} pipeline \\
				\end{tabular}}\\  \cmidrule[0.05em]{1-2}
			
		    	\multicolumn{1}{c}{ \textbf{Model Monitoring (mm)} } &
				\multicolumn{1}{l}{\begin{tabular}[c]{@{}l@{}} 	
				\tabitem  Data deviation\\
				\tabitem  Retrain model when data deviates from original \\
				\end{tabular}}
				\\  \cmidrule[0.05em]{1-2}
				
				\multicolumn{1}{c}{   \textbf{Implementation* (i)} }&			\multicolumn{1}{l}{\begin{tabular}[c]{@{}l@{}} 	
						\tabitem  Facilitate replicability \\
						\tabitem  Facilitate  reproducibility\\
						\tabitem  Docuemntation (experiments and (hyper)parameters)\\
						\tabitem  Resource aware systems  (resource usage)
				\end{tabular}}
				\\  \bottomrule
				\multicolumn{2}{l}{* This stage is not proposed by \cite{amershi2019software}}\\\addlinespace[1pt]
			\end{tabular}
	\end{table}
\end{small}

%% file: table_practices_mr.tex
\begin{small}
	\setlength{\tabcolsep}{2pt}
	\begin{longtable}{M{20pt} m{225pt} m{25pt} m{25pt}} 
		
		\caption{Validated practices for Model Requirement. Including the extra considerations, where  P - Phase category related  notes, C - additional clarifications} \label{tab:pr_mr} \\
		
		\toprule
		\endfirsthead
		
		\multicolumn{4}{c}%
		{{\bfseries \tablename\ \thetable{} -- continued from previous page}} \\
		\\ \hline 
		\endhead
		
		\hline \multicolumn{4}{r}{{Continued on next page}} \\ \hline
		\endfoot
		
		\endlastfoot
	 	\multicolumn{1}{c}{\textbf{ID}}& \multicolumn{1}{c}{\textbf{Description}} & \multicolumn{1}{c}{\begin{tabular}[c]{@{}c@{}}
	 			\textbf{Validated*} \end{tabular} 
 			} & \multicolumn{1}{c}{\begin{tabular}[c]{@{}c@{}}
 				\textbf{Extra**}  \end{tabular} }\\
		\midrule
		\multicolumn{1}{l}{\textbf{\acron[M]{mr}{1}}}& The goal of the model that is going to be trained should be considered when selecting the most appropriate metrics for assessing the model performance. & 4 &-\\ \cmidrule[0.01em]{1-4} \addlinespace[1pt] 
		\multicolumn{1}{l}{\textbf{\acron[M]{mr}{2}}} &  It is important to identify the needs/requirements of model retraining.& 4&-\\ \cmidrule[0.01em]{1-4} \addlinespace[1pt] 
		\multicolumn{1}{l}{\textbf{\acron[M]{mr}{3}}}  & If an ML model is published as a cloud service, specifically, when  a classification/prediction task uses ML cloud-based services, it is important  to define the use case and the model requirements in order to identify how frequently the service should be invoked. & 4&-\\\cmidrule[0.01em]{1-4} \addlinespace[1pt] 
		\multicolumn{1}{l}{\textbf{\acron[M]{mr}{4}} }& When using probabilistic forecasting in a decision system, it is necessary to decouple the probabilistic model optimization from the probability threshold selection. & 3 & P\\
		\bottomrule
		\addlinespace[1pt]
		\multicolumn{3}{l}{* Number of \ac{ML} experts (out of 4) that validated each practice}\\
		\multicolumn{3}{l}{** Extra considerations}
	\end{longtable}
\vspace{-8.2pt}
\end{small}

%% file: table_practices_dc.tex
\begin{small}
	\setlength{\tabcolsep}{2pt}
	\begin{longtable}{M{20pt} m{225pt} m{25pt} m{25pt}} 
		
		\caption{Validated practices for Data Collection, including extra considerations:  P (Phase category comments), C (additional clarifications)} \label{tab:pr_dc} \\
		
		\toprule
		\endfirsthead
		
		\multicolumn{4}{c}%
		{{\bfseries \tablename\ \thetable{} -- continued from previous page}} \\
		\\ \hline 
		\endhead
		
		\hline \multicolumn{4}{r}{{Continued on next page}} \\ \hline
		\endfoot
		
		\endlastfoot
		\multicolumn{1}{c}{\textbf{ID}}& \multicolumn{1}{c}{\textbf{Description}} & \multicolumn{1}{c}{\begin{tabular}[c]{@{}c@{}}
				\textbf{Validated*} \end{tabular}
		}& \multicolumn{1}{c}{\begin{tabular}[c]{@{}c@{}}
			\textbf{Extra**} \end{tabular} }\ \\
		\midrule
		\textbf{\acron[M]{dc}{1}} & The distribution of the training data should reflect the real distribution. & 4& -\\ \cmidrule[0.01em]{1-4} \addlinespace[1pt] 
		\textbf{\acron[M]{dc}{2}}  &If a model is expected to detect ''something" in an image, then representative examples of that ``something" should be present in the training and testing data.& 4&C\\ \cmidrule[0.01em]{1-4} \addlinespace[1pt] 
		\textbf{\acron[M]{dc}{3}} &When measuring a dataset size it should not be done only by referring to storage space but also in terms of rows and columns. & 4 &C\\ \cmidrule[0.01em]{1-4} \addlinespace[1pt] 
		\textbf{\acron[M]{dc}{4}} &If it is required to augment the number of instances in the negative class, preexisting datasets could be used for including more instances in the dataset.& 3 & -\\ \cmidrule[0.01em]{1-4} \addlinespace[1pt] 
		\textbf{\acron[M]{dc}{5}} &The images that are going to be used as instances of the negative class should have some common characteristics with the positive ones. & 3 & -\\ \cmidrule[0.01em]{1-4} \addlinespace[1pt] 
		\textbf{\acron[M]{dc}{6}} & The minimum size of the object that is going to be detected should be present in the data that is going to be used for training the model. & 3 & -\\ \cmidrule[0.01em]{1-4} \addlinespace[1pt] 
		\textbf{\acron[M]{dc}{7}} &The object ROI should have a similar aspect ratio in all the positive images. & 3 & -\\ 
		\bottomrule
		\addlinespace[1pt]
		\multicolumn{3}{l}{* Number of \ac{ML} experts (out of 4) that validated each practice}\\
		\multicolumn{3}{l}{** Extra considerations}
	\end{longtable}
\vspace{-8.2pt}
\end{small}

%% file: section_discussion.tex
\section{Discussion}

This section presents the discussion in three main parts. The first part presents a discussion about topics covered and not covered by the identified practices; we also present a comparison between the topics covered in related work and important aspects that were not covered. Then, we present a discussion about points to note in the practice validation, including a brief discussion about the practices that were not validated by the experts. Finally, we end the section with recommendations for practitioners. 


\subsection{Practices and \ac{ML} stages}

The 127 identified,  validated, and presented practices are distributed in the \ac{ML} pipeline stages, as shown in \figref{fig:stages}.   As it can be seen, most of the practices 
were located in the  \textit{Model Training} (47 practices, 37\%) and \textit{Data Cleaning} (33 practices, 25.98\%) stages.  This large number of practices in both stages could indicate  practitioners' interest in those stages. Specifically, in  \textit{Model Training}, the interest could be related to the importance of the stage in building an \ac{ML} system since  \ac{ML} models are built-in that stage. In addition, to the importance of that stage, according to \cite{anacondainc_2022}, data scientists spent a substantial  percentage of their time  (12\%)   training  a model. 
 
Upon further inspection in the \textit{Model Training} stage, when analyzing the topics covered by the identified practices, we noticed that various topics were covered  (see \tabref{tab:summary}), which are also discussed in the related work, and some others were not discussed. One of the main topics was ``how to split a dataset for training a model''. This topic was also covered in the guidelines and practices presented by  \cite{halilaj2018machine, MichaelLones2021, teschendorff2019avoiding, wujek2016best}. In particular, they emphasize that only a special part of the dataset should be used for training models and another for evaluating them. In addition, they also mentioned the usage of cross-validation (CV) for training and testing the model when having a small amount of data and nested CV when hyper parameter tuning is needed.  Note that some of the literature that presents mismatches or/and guidelines for \ac{ML} does not cover this stage in-depth due to their approach \citep[\eg][]{ horneman2020ai, zinkevich_2021} or the stage was not considered in the study \citep{biderman2020pitfalls}. In addition to that topic, the identified practices also covered ensemble learning, particularly how and when to build a bagging ensemble;  \cite{zinkevich_2021} also discussed this topic, but they only mentioned the different types of ensembles. Moreover,  \cite{wujek2016best} also noted ensemble learning as a way to use the strengths of each model when using it for a classification task.  Otherwise, practices regarding how to train a model to be robust, transfer learning, retraining a model,  and types (\ie batch, min batch, online)  were not discussed in any of the white and gray literature presented in section \ref{sec:related2}.

Regarding the possible interest in the \textit{Data Cleaning} stage, it could be related to the perception of the stage as a challenging one, as mentioned by  \cite{Islam_2019} and \cite{Alshangiti_2019} (see section \ref{sec:related}). This opens a possible necessity for guidelines and practices to overcome  those challenges. In addition, according to  \cite{anacondainc_2022}, in that stage is in which data scientists spend more of their time (39\%) (\ie 22\% in data preparation and 17\% in data cleansing). Note that \citeauthor{anacondainc_2022} exposed the importance of this stage and the human intervention, indicating: ``while data preparation and data cleansing are time-consuming and potentially tedious, automation is not the solution. Instead, having a human in the mix ensures data quality, more accurate results, and provides context for the data.''

When analyzing the \textit{Data Cleaning} stage in detail, it is possible to note that it encompasses a broader set of topics like understanding data, ensuring data quality, transforming data, handling errors, and handling missing data. Some  of these topics are also covered by articles presented in the related work (\ref{sec:related2}). In particular, \cite{biderman2020pitfalls} discussed data quality but by analyzing the heterogeneity of the datasets to avoid bias in the model in the training stage; the practices listed in our study are more related to how to (i) calculate statistics properly, (ii) check errors, (iii) format data.  Moreover, \cite{MichaelLones2021} discussed: guidelines  about the need to take time to understand the data and errors in it; he also emphasized that only the training data should be analyzed and not the whole dataset in order to avoid data leakage;  he also discussed having enough data to train useful models. These aforementioned topics are also covered by the identified practices in our study and by  \cite{wujek2016best}. In addition, \citeauthor{wujek2016best}  also focused on guidelines for standardization, emphasizing that these techniques could affect the results of trained algorithms. They also presented guidelines for handling noisy data such as outliers, encoding numerical variables into bins, and handling missing values. Regarding the last topic, while \citeauthor{wujek2016best} presented the types of missing data briefly and some practices on how to handle missing data, the identified practices regarding that topic are about identification and how to propagate imputation methods in the whole dataset. Some of the identified practices (\ie practices related to transformation and imputation) by this study and the previously mentioned in this paragraph are part of an aspect of data cleaning entitled data tidying.  This aspect of data cleaning  consists of structuring datasets to facilitate analysis, which \citeauthor{Wickham_2014} describes. Conversely to the studies mentioned in the related work, in that stage, the identified practices, in this article,  also covered aspects regarding data anonymization, encoding geographical  and datetime data, and working with handwritten text. This could be because practitioners contributing to \ac{STE} websites work on a variety of disciplines and projects.

The next two stages  \textit{Feature Engineering} and \textit{Implementation},  have  less number of practices than the first two, 11 practices.   Note that the former stage  is considered by some practitioners as part of data preparation. For example, it is not clear in which step is located feature engineering in the survey executed by \cite{anacondainc_2022}; therefore,  it could also be related to the 39\% of  a data scientist time previously mentioned. 

When analyzing  the identified practices at topic granularity in the  \textit{Feature Engineering}  stage,  they  encompass topics such as  aspects to consider while using techniques for feature selection (when, where, how), dimensional reduction, and assumptions of the  feature selection techniques.  In particular,  a general practice that established that feature selection should be done only by taking into account the training data is also mentioned by \cite{MichaelLones2021, teschendorff2019avoiding}. Furthermore, \cite{zinkevich_2021} gave general tips for feature engineering, mentioning that features that are not useful should be eliminated; that is preferable simple features over complex ones; and should be explored features that generalize across features. Finally, more than practices or guidelines, \cite{wujek2016best}  presented which types of dimensional reduction techniques exist (\eg non-negative matrix factorization, PCA). This perspective is complementary to some of the identified practices in our study since those practices mentioned that before executing a dimensional reduction of feature selection technique, the existing techniques for that purpose should be checked.  Beyond the covered topics, when analyzing the scope of those, bear in mind that most of the identified practices are about dimensional reduction. This feature selection process in deep learning models is not always used/applied since layers of deep Neural Networks perform it.

With respect to the \textit{implementation} stage, various topics related to one or more \ac{ML} pipeline stages were covered by the  practices validated in our study. This stage is related more to the \textit{Software Engineering} perspective since it is related to what is needed to implement an \ac{ML} model (\eg resources).  However, those requirements/practices would not exist without \ac{ML}; therefore, we considered them essential to be presented. In particular, according to \cite{horneman2020ai}, it is important to consider all the possible resources (computing, hardware, storage, bandwidth, expertise, and time) that an artificial intelligence 
system requires throughout the system's life.  Note that underestimating the needed resources of a system can cause that at some point, the system crashes due to the lack of them. Another aspect considered by \cite{horneman2020ai} is  about versioning model and data versions for potential needs of recovery, traceability, and decision justification, but they do not provide  guides about how to do it.  

Regarding the \textit{Model Evaluation} stage (8 practices), while the identified practices cover a variety of topics such as using adversarial inputs to ensure robustness, checking bias for NLP tasks, and taking more than only the  performance to evaluate a model,  \citeauthor{MichaelLones2021},  and \citeauthor{biderman2020pitfalls} covered topics like the usage of statistical methods when comparing models. Note that using statistical methods to compare models is a must that has been discussed previously in  \citep{janez_2006}. In addition, \citeauthor{MichaelLones2021}  adds that when comparing the results of a model with other studies, it should be done carefully since the results could be affected by external factors, such as which part of data is used for testing and the range of hyper-parameters tested.

Concerning \textit{Data Collection} (7 practices), the validated practices did not emphasize the process itself of collecting data; however, they focused on the collected data characteristics. This same behavior also occurs in the study conducted by \citeauthor{MichaelLones2021}. One possible explanation for this  is that in many studies and projects, the data is provided by external resources or obtained in publicly/owned  datasets. However, some pitfalls can occur during the collection process. For example, if the right tools are not used to collect data, then the acquired data could have errors or missing data. Another example of a possible error when not following a correct procedure for collecting data is not establishing the minimum standards that the data has to accomplish. For instance, if data standards like mandatory  fields or the data collection frequency are not established from the start, the acquired data could be completely useless.  One practice that could help to collect useful data is mentioned by  \cite{biderman2020pitfalls}, which indicates that a good approach to collect data would be to design a collection strategy with a hypothesis in mind. 

In the \textit{Model requirement} stage, the validated practices focus on the four topics depicted in \tabref{tab:summary}. For instance,  some practices stand out the importance of identifying model assumptions and recognizing the goal of the \ac{ML} task to select the best possible model to accomplish a particular task. Both  topics were also discussed by 
 \citeauthor{biderman2020pitfalls, MichaelLones2021, halilaj2018machine}  in their studies. 
Furthermore, \cite{horneman2020ai}  and \cite{zinkevich_2021} emphasize that not  all the problems could be solved with \textit{AI} or \textit{\ac{ML}}, and before going into further stages in the \ac{ML} pipeline, it should be checked that the problem could be solved with that kind of solutions.  Additionally,  one of the validated practices also describes the importance of the selection of the metric used for evaluating a model. This practice is also mentioned by the studies of \cite{halilaj2018machine, MichaelLones2021, zinkevich_2021}. Unlikely the studies in section  \ref{sec:related2}, the validated practices do cover topics related to probabilistic forecasting methods, and cloud services are discussed. 

The less discussed stages, \ie the stages with the least number of practices, were \textit{Data Labeling}, \textit{Model Deployment}, and  \textit{Model Monitoring}, with only two practices per stage.  The low number of identified practices in   \textit{Data Labeling} could be associated with the fact that in some projects, the collected/obtained data is already labeled; therefore, the  efforts are focused on other stages. Concerning the topics covered by the identified practices in the  \textit{Data Labeling} stage, they are mainly about the scalability of the process and parameterizing regions of interest for object detection, leaving aside other topics, like  the expertise of  taggers or  labeling noise  (\ie possible errors made in the process) mentioned by \cite{biderman2020pitfalls}.  The presence of these topics in the study made by  \citeauthor{biderman2020pitfalls} could be related to its focus (\ie they are focused on \ac{ML} research).

Regarding the low number of practices in the  \textit{Model Deployment} and \textit{Model Monitoring} stages, a potential explanation is that these stages are more related to the ``operational''  side, especially the \textit{Model Monitoring} stage, as mentioned by \cite {LewisGrace2021WAIN}.  This is supported by the fact that data scientists spent less time deploying models than in other tasks (\ie model training, model selection, data visualization, reporting and presentation, data cleansing, and data preparation).  Additionally, model monitoring was not even considered as a possible task/stage in which data scientists spent their time in the study done by \citeauthor{anacondainc_2022}, which  could cause mismatches between the operational side and data scientists, like the ones mentioned by \cite{LewisGrace2021WAIN}. 

When analyzing the few  validated practices in the  \textit{Model Deployment} stage, we see an opportunity for future work. Especially when analyzing the study made by \cite{LewisGrace2021WAIN}, in which  the majority of the identified mismatches refer to incorrect assumptions about the trained model when data scientists transfer it to software engineers.  Moreover,  \cite{Islam_2019}  identified that one of the most challenging stages was   \textit{Model Deployment}. 

The validated  \textit{Model Monitoring} practices  are about data deviation and re-training as a contingency action.  Regarding possible contingencies, a practice was proposed by  \cite{wujek2016best}, in which they describe  a Champion-challenger testing deployment technique 
The other study that also discussed guidelines for that stage was made by \cite{zinkevich_2021}. In particular, they mentioned the importance of recognizing the necessities of the freshness of a model, which help in the understanding of the monitoring priorities. In addition, they emphasize that in the monitoring process, aspects such as ``silent failures'' (\ie errors that normally do not raise an error or crash a system) should also be  checked. 

One important topic not discussed by the validated practices is ethics, which is mentioned by \citep{horneman2020ai}. The authors  discussed that ethics should be considered as a software design consideration and a  policy concern, and  every aspect of the system should be studied for potential ethical issues. In particular, they mentioned that data collection often raises questions about privacy, and how the final system will be used is also an ethical concern. Regarding this topic, we encourage the readers to read the complete framework proposed by IBM \citep{ibm_ethics}. This framework establishes five principles (\ie explainability, fairness, robustness, transparency, and privacy) that will help to establish an ethical foundation for building and using artificial intelligence systems. Regarding fairness in artificial intelligence, some works have been done in which bias  is analyzed  to understand  how it is done and what to do \citep[\eg][]{chakraborty2021bias}.  Moreover, fairness in software  has also been studied \citep[\eg][]{brun2018software, soremekun2022software}. This point of view of fairness is important since, as we previously mentioned, \ac{ML} has been adopted in multiple \ac{SE} tasks and has enabled systems.

\subsection{Expert Opinions}

While the experts often agreed with each other and all validated 70 of the practices, there were also 58 practices to which only three experts agreed and 30 practices to which less than three experts agreed. Many cases where one expert disagreed were not conflict with the practice itself, but rather the placement of the practice within the taxonomy (\eg \acron{mr}{4}, \acron{mt}{27}, \acron{mt}{28}, \acron{mt}{42}, \acron{mt}{43}), regarding wording within the practice or clarifications (\eg \acron{dcl}{11}, \acron{mt}{16}, \acron{mt}{30}), through the specification of condictions under which something should be considered a best practice (\eg \acron{dcl}{24}, \acron{mt}{21}),  because of details the expert would have also expected within the practice (\eg \acron{dcl}{26}), the question if a practice is sufficiently specific (\eg \acron{fe}{6}), that a practice is good but requires expertise that is usually lacking (\eg \acron{fe}{8}). There is only one case in which an expert disagreed and left a comment that indicates a strong and active disagreement (\eg mt-45). Thus, the expert opinions are, in general, well-aligned with the validated practices. 

For the practices that  were not validated, the main cause was that the experts marked them as UNCLEAR, \ie that the experts did not understand what the practice was about. We believe there are two likely reasons for this. First, the wording of the practice may be bad, meaning that it is ambiguous, unspecific, or uses unknown terminology. Second, the practices may sometimes only be understandable in the context of the STE posts where they were suggested, which would indicate that they are not sufficiently general to be a best practice, as they are rather context-specific solutions. 

Another notable aspect of the expert opinions is that they agreed to contradicting practices. Practice \acron{me}{2} states that the test data should only be used to evaluate the best model, and practice \acron{me}{4} states that the best model should be selected on the test data. Three of the four experts agreed with both practices. We believe that this is an indication that the practices need to be extended with the scenario in which they are suitable. For example, when developing an application with an AI component, the test data should certainly only be used once. However, when talking about machine learning benchmarks like ImageNet~\citep{imagenet_2009} or SuperGLUE~\citep{wang2019superglue} the test data is used by many research groups to select the best model. 

\subsection{Lessons Learned}

One notable aspect of our study is that we did not specify what the best practices were, but instead, we relied on \ac{STE} posts to identify what the users of those websites considered as best practices. Therefore, the practices reported in this paper are active recommendations that result on problems reported by developers. This led to the problem, that some practices were similar, but covered different nuances, leading to some redudancy within our catalog of practices (\eg \acron{dcl}{9} and \acron{dcl}{10}) and even contradictions, as we discussed above. While we further harmonized the practices as even removed contradictions as part of the manual coding to create the taxonomy, we decided against this. Instead, in our study we relied on the experts judgement. However, we believe that this is an important consideration for studies based on data from \ac{CQ+A} sites: not every answer is good and not every answer is generalizable beyond the immediate context of a question. This possibly affects an study that analyzes the content of \ac{CQ+A} sites with the goal to understand discussion around a certain issue. For researchers, we recommend to carefully decide how the quality of answers is evaluated, to understand how interactions between related answers affect the planned analysis.
\vspace{-5pt}
\subsection{Recommendations for Practitioners}

The list of practices from \ac{CQ+A} sites we collected provides a good resource for practitioners to enhance their process of developing applications with machine learning components. However, our analysis shows that not everything that is recommended by the crowd can be used without analyzing the implementation context, as our four experts only had unanimously agreement for less than half of the overall practices. Consequently, developers should consider recommended best practices --- from  our catalog,  answers on \ac{CQ+A} sites, and any other sources--- always carefully within the context of individual projects to determine what are really the best practices for the context.  


%% file: section_threats.tex
\section{Threats to Validity}
\label{sec:threats}

We report the threats to the validity of our work following the classification by \cite{Cook1979} suggested for software engineering by \cite{Wohlin2012}. Additionally, we discuss the reliability as  \cite{Runeson2009} indicate.

\textbf{Threats to construct validity}  refer to inappropriate instruments or data collection 	methods to evaluate hypothetical constructs. For instance, the method used for selecting a reported practice as a best practice could be a practice in the context of this study. However, we mitigated that threat by relying on the judgment of four machine learning experts. In addition, the selection of the post representing questions and answers related to machine learning could be a threat. In this study, we retrieved 121 pairs of questions and answers from 14 \ac{STE} pages, and we can not claim that those 121 pairs are the whole population of questions and answers mentioning machine learning best practices. We relied on selecting papers with the ``machine learning'' tag to reduce the number of false positives, and because the  tags-based selection is a common practice in papers that analyze posts from questions and answers systems.

\textbf{Threats to internal validity}  relate to  when causal relations are examined, and there are confounding factors or decisions  that do not allow the measurement of the relationship between the results and an intervention. In our case, this is more an exploratory qualitative study, and no intervention was done like in the case of an experiment with control and treatment groups. However, a potential threat could be  if the post-selection was based on the scores of the \ac{STE} contributors. We avoided that threat because the posts were not selected with that strategy. We did filter the posts based on the score of the post, to make sure that the posts are considered useful by the community; this type of selection does not introduce a threat to internal validity, and filtering based on posts' scores is a widely used  strategy followed by studies that analyze posts from question and answer websites.


\textbf{Threats to external validity}  refer to  the generalization of results. In this study, we can not claim that our results generalize to the whole population of machine learning best practices.  Although we have incorporated  127 \ac{ML} practices discussed by practitioners in the different pipeline stages covering a range of topics/themes, we can not claim our study covers all the  multiple  \ac{ML} areas in which practices are discussed. In this regard, our study can be improved by including more data sources in which more \ac{ML} subjects are discussed. In addition, the 127 presented practices are not a ``gold standard'' since, in the \ac{ML} field, several exceptions to these practices could happen due to the multiple variables present in an \ac{ML} project (\eg data distribution,  type of models,  type of tasks). That is why from the beginning of our study, we claimed the multiple possible exceptions to the practices.  

In addition to the aforementioned threats, we have some additional \textbf{internal} and \textbf{conclusion} threats reported as threats to reliability, as follows.
\textbf{Reliability}  focuses on to what extent the data and the analysis are dependent on the specific researchers. In our study, this relates to (i) \textit{Subjectivity in the \ac{STE} websites selection}: through manual analysis of the title and description, we identified possible \ac{STE} websites in which we could find posts related to \ac{ML}. In order to mitigate this subjectivity bias in the selection page process, the pages were selected by a group of four people and then discussed to agree to the final 14 lists of pages.  In addition, as discussed previously, for the construct threats, we can not claim our sample of 121 pairs of questions and answers contains the whole population of posts in STE related to machine learning best practices.

Another threat related to  reliability is  (ii)  \textit{Subjectivity in the  post tagging}. To mitigate this bias, we conducted the tagging process with a group of three people with different backgrounds, in which at  least two people tagged each post. The conflicts were solved by each pair of taggers in meetings; in each meeting, each tagger explained their arguments to reach an agreement. 
Another potential threat to reliability is (iii)   \textit{subjectivity in the \ac{STE} practice classification of subcategories and topics due to an open-coding scheme;} to mitigate this bias in the validation process, the \ac{ML} experts had the option of objecting about the category and topic a practice was assigned to; then, throughout the article, we presented all the cases in which at least one expert objected. Moreover, we analyzed their comments, and in the cases that they were valid, we moved the practice to a new categorization. Finally, another threat we discuss is (iv) \textit{Subjectivity in identifying best practices}; in the tagging process, the taggers did not evaluate if the identified  practice was 
a good one because this validation (\ie the process to identify if a practice was a best one or not ) was made by four experts with different \ac{ML} backgrounds.

%% file: section_conclusion.tex

\section{Conclusion \& Future Work}

Machine learning has been widely used in domains such as software engineering, in which it provides tremendous support  for the automation of highly important tasks (\eg code recommendation, automated patching, bug triage). However, its usage as a black-box or off-the-shelf component could lead to pitfalls, unexpected results, or mismatches, especially when the \ac{ML} methods/techniques are not used properly. In fact, some previous works depicted in Section \ref{sec:related}  have reported some of those pitfalls and mismatches. Moreover, there is a plethora of works describing how to use \ac{ML} from different perspectives, which makes it hard to find/browse a comprehensive source that could help software engineers (and researchers/practitioners from other fields)  to understand \ac{ML} best practices. There are some initial efforts in that direction, and with this paper, we complement existing work and contribute to the goal of having a comprehensive list of best practices by building a taxonomy of  \ac{ML} best practices, firstly reported by practitioners (via STE discussions) and lately validated by four \ac{ML} experts. 

We expect the practices presented in this paper could help software engineers (researchers and practitioners)  to understand how to use ML in a more informed way. We do not claim this is a complete ``handbook'' of best practices. However, we present a list of 127 practices covering all the stages from the \ac{ML} pipeline proposed by~\cite{amershi2019software} and one additional stage not included there (\ie implementation).  These large number of practices and the plethora of existing works around \ac{ML} suggest that the \ac{ML} body of knowledge is continuously growing,  which increases the complexity of using \ac{ML} properly. Therefore, more work should be devoted to improving the ways how researchers and practitioners can access that body of knowledge. For instance,   taking advantage of recent progress on Robotic Process Automation (RPA) and recommender systems, we could create bots and recommenders that would  support software engineers to (i) select \ac{ML} best practices under specific contexts and conditions,  (ii) design experiments that involve \ac{ML} components, and (iii) implement systems that include \ac{ML} components. Moreover, future work could be devoted to analyzing  \ac{ML}  practices conducted by software engineers (researchers and practitioners)  to validate their correctness, report pitfalls and mismatches, and  continue creating a handbook of best practices.